\documentclass[preprint,3p]{elsarticle}

\usepackage{amsmath,
	    amsfonts,
	    amstext,
	    amssymb,
	    mathrsfs,
	    stmaryrd}
\usepackage{latexsym}
\usepackage{graphicx}
\usepackage{epsf}

\def\ms#1{\null\ifmmode\mathord{\mathcode`-="702D\it #1\mathcode`\-="2200}%
	\else$\mathord{\mathcode`-="702D\it #1\mathcode`\-="2200}$\fi}

\newcommand{\cws}[2]
	{\\ \centerline{$#2$} \\[-#1pt]}

\newlength{\spacelen}

\newcommand{\bibtrick}[1]
	{}

\newcommand{\calb}
        {\mathcal{B}}

\newcommand{\calc}
        {\mathcal{C}}

\newcommand{\calcc}
        {\mathcal{CC}}

\newcommand{\calccc}
        {\mathcal{CCC}}

\newcommand{\cald}
        {\mathcal{D}}

\newcommand{\calfcc}
        {\mathcal{FCC}}

\newcommand{\calftcc}
        {\mathcal{FTCC}}

\newcommand{\calg}
        {\mathcal{G}}

\newcommand{\call}
        {\mathcal{L}}

\newcommand{\calrcc}
        {\mathcal{RCC}}

\newcommand{\calrtcc}
        {\mathcal{RTCC}}

\newcommand{\cals}
        {\mathcal{S}}

\newcommand{\calsc}
        {\mathcal{SC}}

\newcommand{\calscc}
        {\mathcal{SCC}}

\newcommand{\calt}
        {\mathcal{T}}

\newcommand{\calz}
        {\mathcal{Z}}

\newcommand{\natns}
	{\mathbb{N}}

\newcommand{\realns}
	{\mathbb{R}}

\newcommand{\step}[2]
        {\, {\auxstep\limits^{#1}}_{#2} \,}
\newcommand{\auxstep}
	{\mathop{-\hspace{-0.15cm}\mapsto}}

\newcommand{\arrow}[2]
        {\, {\auxarrow\limits^{#1}}_{#2} \,}
\newcommand{\auxarrow}
	{\mathop{\longrightarrow}}

\newcommand{\sbis}[1]
	{\sim_{#1}}

\newcommand{\pre}[1]
        {\sqsubseteq_{#1}}

\newcommand{\fullbox}
	{{\mbox{}\nolinebreak\hfill{$\rule{2mm}{2mm}$}}}

\newtheorem{new_theorem}
	{Theorem}[section]

\newtheorem{new_definition}
	[new_theorem]{Definition}

\newtheorem{new_remark}
	[new_theorem]{Remark}

\newtheorem{new_example}
	[new_theorem]{Example}

\newtheorem{new_lemma}
	[new_theorem]{Lemma}

\newtheorem{new_proposition}
	[new_theorem]{Proposition}

\newtheorem{new_corollary}
	[new_theorem]{Corollary}

\newenvironment{definition}
	{\begin{new_definition}\rm}
	{\end{new_definition}}

\newenvironment{theorem}
	{\begin{new_theorem}\rm}
	{\end{new_theorem}}

\newenvironment{proof}
	{\medskip\noindent{\bf Proof}$\ $}
	{}

\newif\ifwithtikz
\withtikzfalse
\ifwithtikz
\usepackage{tikz}
\usetikzlibrary{external}
\fi

\journal{}

\begin{document}

\begin{frontmatter}

\title{A Companion of ``Relating Strong Behavioral Equivalences \\
       for Processes with Nondeterminism and Probabilities''}

\author[uniurb]{Marco Bernardo}
\author[imtlu]{Rocco De Nicola}
\author[unifi]{Michele Loreti}

\address[uniurb]{Dipartimento di Scienze di Base e Fondamenti -- Universit\`a di Urbino -- Italy}
\address[imtlu]{IMT -- Institute for Advanced Studies Lucca -- Italy}
\address[unifi]{Dipartimento di Statistica, Informatica, Applicazioni -- Universit\`a di Firenze -- Italy}


\begin{abstract}
In the paper ``Relating Strong Behavioral Equivalences for Processes with Nondeterminism and Probabilities''
to appear in Theoretical Computer Science, we present a comparison of behavioral equivalences for
nondeterministic and probabilistic processes. In particular, we consider strong trace, failure, testing, and
bisimulation equivalences. For each of these groups of equivalences, we examine the discriminating power of
three variants stemming from three approaches that differ for the way probabilities of events are compared
when nondeterministic choices are resolved via deterministic schedulers. The established relationships are
summarized in a so-called spectrum. However, the equivalences we consider in that paper are only a small
subset of those considered in the original spectrum of equivalences for nondeterministic systems introduced
by Rob van Glabbeek. In this companion paper, we enlarge the spectrum by considering variants of trace
equivalences (completed-trace equivalences), additional decorated-trace equivalences (failure-trace,
readiness, and ready-trace equivalences), and variants of bisimulation equivalences (kernels of simulation,
completed-simulation, failure-simulation, and ready-simulation preorders). Moreover, we study how the
spectrum changes when randomized schedulers are used instead of deterministic ones. 
\end{abstract}


\end{frontmatter}

%
%
\section{Introduction}\label{sec:intro}
%
%

\noindent
In~\cite{BDL13TCS}, a systematic account of the main known probabilistic equivalences for nondeterministic
\emph{and} probabilistic systems has been presented by defining them over an extension of the LTS model
combining nondeterminism and probability that we call NPLTS, in which every action-labeled transition goes
from a source state to a probability distribution over target states rather than to a single target
state~\cite{LS91,Seg95a}. Schedulers, which can be viewed as external entities that select the next action
to perform, are used to resolve nondeterminism~\cite{Seg95a}. By ``playing'' with schedulers, a number of
possibilities for defining behavioral equivalences over NPLTS models emerge. We concentrated on three
approaches that differ for the way probabilities of events are compared when nondeterministic choices are
resolved via schedulers:

	\begin{description}

\item[1. Fully Matching Resolutions] Two resolutions are compared with respect to the probability
distributions of all considered events. 

\item[2. Partially Matching Resolutions] The probabilities of the set of events of a resolution are required
to be individually matched by the probabilities of the same events in possibly different resolutions.

\item[3. Max-Min-Matching Resolution Sets] Only the extremal probabilities of each event stemming from the
different resolutions are compared.

	\end{description}

In~\cite{BDL13TCS}, we have studied the relationships among the probabilistic variants of the main
equivalences for nondeterministic systems that stem from the three approaches outlined above. 

\begin{figure}[tp]

\centerline{\includegraphics{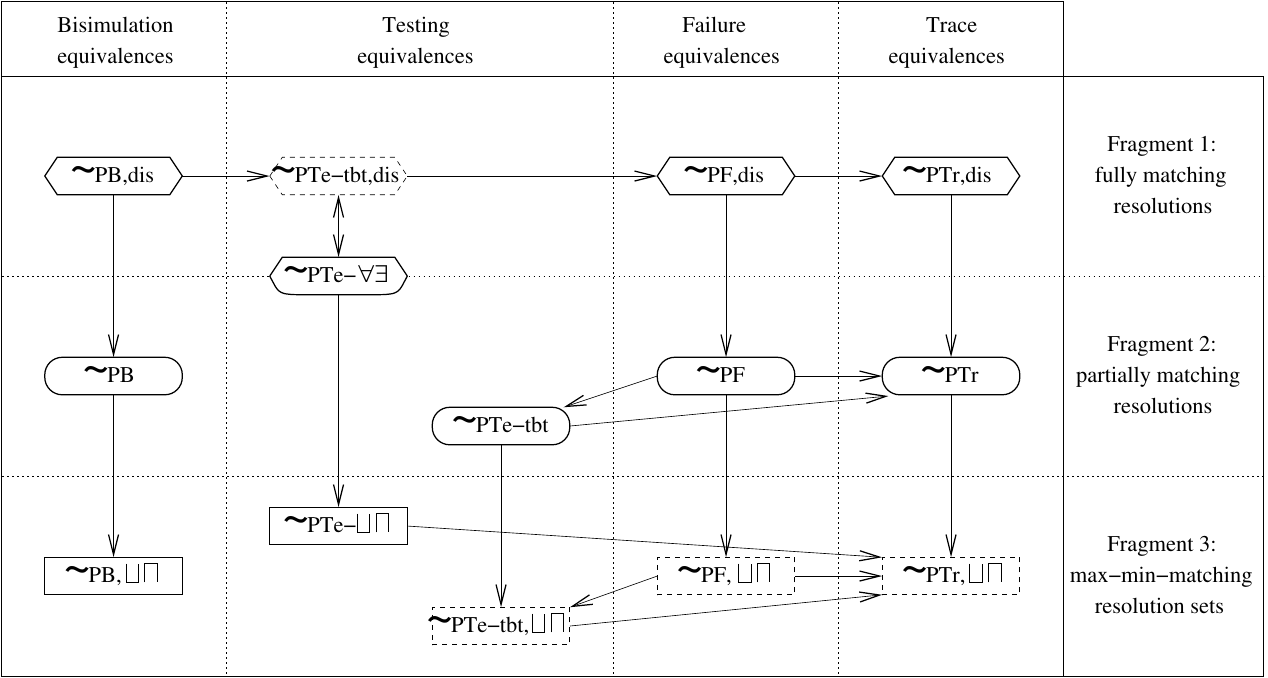}}
\caption{Reduced spectrum of strong behavioral equivalences for NPLTS models (deterministic schedulers)}
\label{fig:spectrum_reduced}

	\end{figure}

We have proposed and analyzed three variants of trace, testing, failure, and bisimulation equivalences for
NPLTS models. Their relationships are summarized in Fig.~\ref{fig:spectrum_reduced}. In the spectrum, the
absence of (chains of) arrows represents incomparability, bidirectional arrows connecting boxes indicate
coincidence, and ordinary arrows stand for the strictly-more-discriminating-than relation. Continuous
hexagonal boxes contain well known equivalences that compare probability distributions of all
equivalence-specific events. In contrast, continuous rounded boxes contain more recent equivalences
assigning a weaker role to schedulers that compare separately the probabilities of individual
equivalence-specific events. Continuous rectangular boxes instead contain old and new equivalences based on
extremal probabilities. The only hybrid box is the one containing $\sbis{\textrm{PTe-}\forall\exists}$, as
this equivalence is half way between the first two definitional approaches. Dashed boxes contain
equivalences that we have introduced to better assess the different impact of the approaches themselves.

In this companion paper, following~\cite{Gla01} we enlarge the spectrum examined in~\cite{BDL13TCS} by
additionally considering variants of trace equivalences (completed-trace equivalences), of decorated-trace
equivalences (failure-trace, readiness, and ready-trace equivalences), and of bisimulation equivalences
(kernels of simulation, completed-simulation, failure-simulation, and ready-simulation preorders). Finally,
we show how the spectrum changes when using randomized schedulers in place of deterministic ones.

We refer the reader to~\cite{BDL13TCS} for motivations and for the description the three approaches to
equivalence definition based on schedulers. Only to guarantee readability, we repeat here the background
section of~\cite{BDL13TCS} that introduces the necessary terminology about NPLTS models and schedulers.

%
%
\section{Nondeterministic and Probabilistic Processes}\label{sec:nplts}
%
%

\noindent
Processes combining nondeterminism and probability are typically described by means of extensions of the LTS
model, in which every action-labeled transition goes from a source state to a \emph{probability distribution
over target states} rather than to a single target state. They are essentially Markov decision
processes~\cite{Der70} and are representative of a number of slightly different probabilistic computational
models including internal nondeterminism such as, e.g., concurrent Markov chains~\cite{Var85}, alternating
probabilistic models~\cite{HJ90,YL92,PLS00}, probabilistic automata in the sense of~\cite{Seg95a}, and the
denotational probabilistic models in~\cite{JSM97} (see~\cite{SD04} for an overview). We formalize them as a
variant of simple probabilistic automata~\cite{Seg95a}.

	\begin{definition}\label{def:nplts}

A \emph{nondeterministic and probabilistic labeled transition system}, NPLTS for short, is a triple $(S, A,
\! \arrow{}{} \!)$ where:

		\begin{itemize}

\item $S$ is an at most countable set of states.

\item $A$ is a countable set of transition-labeling actions.

\item $\! \arrow{}{} \! \subseteq S \times A \times \ms{Distr}(S)$ is a transition relation, where
$\ms{Distr}(S)$ is the set of discrete probability distributions over $S$.
\fullbox

		\end{itemize}

	\end{definition}

A transition $(s, a, \cald)$ is written $s \arrow{a}{} \cald$. We say that $s' \in S$ is not reachable from
$s$ via that $a$-transition if $\cald(s') = 0$, otherwise we say that it is reachable with probability $p =
\cald(s')$. The reachable states form the support of $\cald$, i.e., $\ms{supp}(\cald) = \{ s' \in S \mid
\cald(s') > 0 \}$. We write $s \arrow{a}{} \!$ to indicate that $s$ has an $a$-transition. The choice among
all the transitions departing from $s$ is external and nondeterministic, while the choice of the target
state for a specific transition is internal and probabilistic. An NPLTS represents (i)~a \emph{fully
nondeterministic process} when every transition leads to a distribution that concentrates all the
probability mass into a single target state or (ii) a \emph{fully probabilistic process} when every state
has at most one outgoing transition.

\begin{figure}[tp]

\input{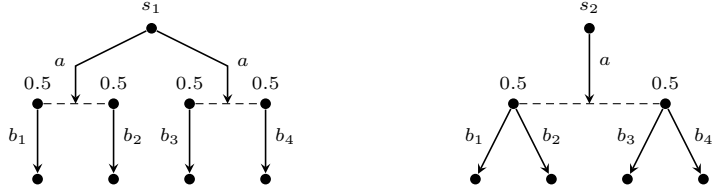}
\caption{Graphical representation of NPLTS models: two examples}
\label{fig:nplts_example}

	\end{figure}

An NPLTS can be depicted as a directed graph-like structure in which vertices represent states and
action-labeled edges represent action-labeled transitions. Given a transition $s \arrow{a}{} \cald$, the
corresponding \linebreak $a$-labeled edge goes from the vertex representing state $s$ to a set of vertices
linked by a dashed line, each of which represents a state $s' \in \ms{supp}(\cald)$ and is labeled with
$\cald(s')$ -- label omitted if $\cald(s') = 1$. Figure~\ref{fig:nplts_example} shows two NPLTS models: the
one on the left mixes internal nondeterminism and probability, while the one on the right does not.

In this setting, a computation is a sequence of state-to-state steps, each denoted by $s \step{a}{} s'$ and
derived from a state-to-distribution transition $s \arrow{a}{} \cald$.

	\begin{definition}\label{def:computation}

Let $\call = (S, A, \! \arrow{}{} \!)$ be an NPLTS and $s, s' \in S$. We say that:
\cws{0}{c \: \equiv \: s_{0} \step{a_{1}}{} s_{1} \step{a_{2}}{} s_{2} \dots s_{n - 1} \step{a_{n}}{} s_{n}}
is a \emph{computation} of $\call$ of length $n$ from $s = s_{0}$ to $s' = s_{n}$ iff for all $i = 1, \dots,
n$ there exists a transition $s_{i - 1} \arrow{a_{i}}{} \cald_{i}$ such that $s_{i} \in
\ms{supp}(\cald_{i})$, with $\cald_{i}(s_{i})$ being the execution probability of step $s_{i - 1}
\step{a_{i}}{} s_{i}$ conditioned on the selection of transition $s_{i - 1} \arrow{a_{i}}{} \cald_{i}$ of
$\call$ at state~$s_{i - 1}$. We say that $c$ is \emph{maximal} iff it is not a proper prefix of any other
computation. We denote by $\ms{first}(c)$ and $\ms{last}(c)$ the initial state and the final state of $c$,
respectively, and by $\calc_{\rm fin}(s)$ the set of finite-length computations from $s$.
\fullbox

	\end{definition}

A resolution of a state $s$ of an NPLTS $\call$ is the result of a possible way of resolving nondeterminism
starting from $s$. A resolution is a tree-like structure whose branching points represent probabilistic
choices. This is obtained by unfolding from $s$ the graph structure underlying~$\call$ and by selecting at
each state a single transition of~$\call$ (\emph{deterministic scheduler}) or a convex combination of
equally labeled transitions of~$\call$ (\emph{randomized scheduler}) among all the outgoing transitions of
that state. Below, we introduce the notion of resolution arising from a deterministic scheduler as a fully
probabilistic NPLTS (randomized schedulers are deferred to Sect.~\ref{sec:det_vs_rand_sched}). Notice that,
when $\call$ is fully nondeterministic, resolutions boil down to computations.

	\begin{definition}\label{def:resolution_det}

Let $\call = (S, A, \! \arrow{}{} \!)$ be an NPLTS and $s \in S$. We say that an NPLTS $\calz = (Z, A, \!
\arrow{}{\calz} \!)$ is a \emph{resolution} of~$s$ obtained \emph{via a deterministic scheduler} iff there
exists a state correspondence function $\ms{corr}_{\calz} : Z \rightarrow S$ such that $s =
\ms{corr}_{\calz}(z_{s})$, for some $z_{s} \in Z$, and for all $z \in Z$ it holds that:

		\begin{itemize}

\item If $z \arrow{a}{\calz} \cald$, then $\ms{corr}_{\calz}(z) \arrow{a}{} \cald'$ with $\cald(z') =
\cald'(\ms{corr}_{\calz}(z'))$ for all $z' \in Z$.

\item If $z \arrow{a_{1}}{\calz} \cald_{1}$ and $z \arrow{a_{2}}{\calz} \cald_{2}$, then $a_{1} = a_{2}$ and
$\cald_{1} = \cald_{2}$.

		\end{itemize}

\noindent
We say that $\calz$ is \emph{maximal} iff it cannot be further extended in accordance with the graph
structure of $\call$ and the constraints above. We denote by $\ms{Res}(s)$ the set of resolutions of~$s$
obtained via a deterministic scheduler and by $\ms{Res}_{\rm max}(s)$ the set of maximal resolutions of~$s$
obtained via a deterministic scheduler. Moreover, we attach subscript $\alpha \in A^{*}$ to those two sets
when we restrict attention to resolutions that have no maximal computations corresponding to proper prefixes
of $\alpha$-computations of~$\call$.
\fullbox

	\end{definition}

Since $\calz \in \ms{Res}(s)$ is fully probabilistic, the probability $\ms{prob}(c)$ of executing $c \in
\calc_{\rm fin}(z_{s})$ can be defined as the product of the (no longer conditional) execution probabilities
of the individual steps of $c$, with $\ms{prob}(c)$ being always equal to $1$ if $\call$ is fully
nondeterministic. This notion is lifted to $C \subseteq \calc_{\rm fin}(z_{s})$ by letting $\ms{prob}(C) =
\sum_{c \in C} \ms{prob}(c)$ whenever none of the computations in $C$ is a proper prefix of one of the
others.

%
%
\section{A Full Spectrum of Strong Behavioral Equivalences}\label{sec:spectrum}
%
%

\noindent
In this section, following~\cite{Gla01} we enlarge the spectrum by additionally considering variants of
trace equivalences (completed-trace equivalences), further decorated-trace equivalences (failure-trace,
readiness, and ready-trace equivalences), and variants of bisimulation equivalences (kernels of simulation,
completed-simulation, failure-simulation, and ready-simulation preorders). Finally, we show how the spectrum
changes when using randomized schedulers in place of deterministic ones.

%
\subsection{Completed-Trace Equivalences}\label{sec:compl_trace_equiv}
%

\noindent
A variant of trace equivalence that additionally considers completed computations was introduced in the
literature of fully nondeterministic models in order to equip trace equivalence with deadlock sensitivity.
Given an NPLTS $\call = (S, A, \! \arrow{}{} \!)$, $s \in S$, $\calz \in \ms{Res}(s)$, and $\alpha \in
A^{*}$, we recall that $\calccc(z_{s}, \alpha)$ denotes the set of \emph{completed} $\alpha$-compatible
computations from $z_{s}$. In other words, each of these computations $c$ belongs to the set $\calcc(z_{s},
\alpha)$ of $\alpha$-compatible computations from $z_{s}$ and is such that $\ms{corr}_{\calz}(\ms{last}(c))$
has no outgoing transitions in~$\call$.

	\begin{definition}\label{def:pctrdis} 

(\emph{Probabilistic completed-trace-distribution equivalence} -- $\sbis{\rm PCTr,dis}$) \\
$s_{1} \sbis{\rm PCTr,dis} s_{2}$ iff for each $\calz_{1} \in \ms{Res}(s_{1})$ there exist $\calz_{2},
\calz'_{2} \in \ms{Res}(s_{2})$ such that \underline{for all $\alpha \in A^{*}$}:
\cws{0}{\begin{array}{rcl}
\ms{prob}(\calcc(z_{s_{1}}, \alpha)) & \!\!\! = \!\!\! & \ms{prob}(\calcc(z_{s_{2}}, \alpha)) \\
\ms{prob}(\calccc(z_{s_{1}}, \alpha)) & \!\!\! = \!\!\! & \ms{prob}(\calccc(z'_{s_{2}}, \alpha)) \\
\end{array}}
and symmetrically for each $\calz_{2} \in \ms{Res}(s_{2})$.
\fullbox

	\end{definition}

	\begin{definition}\label{def:pctr}

(\emph{Probabilistic completed-trace equivalence} -- $\sbis{\rm PCTr}$) \\
$s_{1} \sbis{\rm PCTr} s_{2}$ iff \underline{for all $\alpha \in A^{*}$} it holds that for each $\calz_{1}
\in \ms{Res}(s_{1})$ there exist $\calz_{2}, \calz'_{2} \in \ms{Res}(s_{2})$ such that:
\cws{0}{\begin{array}{rcl}
\ms{prob}(\calcc(z_{s_{1}}, \alpha)) & \!\!\! = \!\!\! & \ms{prob}(\calcc(z_{s_{2}}, \alpha)) \\
\ms{prob}(\calccc(z_{s_{1}}, \alpha)) & \!\!\! = \!\!\! & \ms{prob}(\calccc(z'_{s_{2}}, \alpha)) \\
\end{array}}
and symmetrically for each $\calz_{2} \in \ms{Res}(s_{2})$.
\fullbox

	\end{definition}

	\begin{definition}\label{def:pctrsupinf}

(\emph{Probabilistic $\sqcup\sqcap$-completed-trace equivalence} -- $\sbis{\rm PCTr,\sqcup\sqcap}$) \\
$s_{1} \sbis{\rm PCTr,\sqcup\sqcap} s_{2}$ iff for all $\alpha \in A^{*}$:
\cws{0}{\begin{array}{rcl}
\bigsqcup\limits_{\calz_{1} \in \ms{Res}_{\alpha}(s_{1})} \ms{prob}(\calcc(z_{s_{1}}, \alpha)) & \!\!\! =
\!\!\! & \bigsqcup\limits_{\calz_{2} \in \ms{Res}_{\alpha}(s_{2})} \ms{prob}(\calcc(z_{s_{2}}, \alpha))
\\[0.4cm]
\bigsqcap\limits_{\calz_{1} \in \ms{Res}_{\alpha}(s_{1})} \ms{prob}(\calcc(z_{s_{1}}, \alpha)) & \!\!\! =
\!\!\! & \bigsqcap\limits_{\calz_{2} \in \ms{Res}_{\alpha}(s_{2})} \ms{prob}(\calcc(z_{s_{2}}, \alpha)) \\
\end{array}}
and:
\cws{10}{\begin{array}{rcl}
\bigsqcup\limits_{\calz_{1} \in \ms{Res}_{\alpha}(s_{1})} \ms{prob}(\calccc(z_{s_{1}}, \alpha)) & \!\!\! =
\!\!\! & \bigsqcup\limits_{\calz_{2} \in \ms{Res}_{\alpha}(s_{2})} \ms{prob}(\calccc(z_{s_{2}}, \alpha))
\\[0.4cm]
\bigsqcap\limits_{\calz_{1} \in \ms{Res}_{\alpha}(s_{1})} \ms{prob}(\calccc(z_{s_{1}}, \alpha)) & \!\!\! =
\!\!\! & \bigsqcap\limits_{\calz_{2} \in \ms{Res}_{\alpha}(s_{2})} \ms{prob}(\calccc(z_{s_{2}}, \alpha)) \\
\end{array}}
\fullbox

	\end{definition}

We now investigate the relationships of the three completed-trace equivalences among themselves and with the
various equivalences defined in~\cite{BDL13TCS}. As in the fully nondeterministic spectrum~\cite{Gla01},
completed-trace semantics is comprised between failure semantics and trace semantics. This holds in
particular for the completed-trace equivalence based on fully matching resolutions, although completed-trace
semantics coincides with trace semantics in the fully probabilistic spectrum~\cite{JS90,HT92}.

	\begin{theorem}\label{thm:compl_trace_results}

It holds that:

		\begin{enumerate}

\item $\sbis{\rm PCTr,dis} \: \subseteq \: \sbis{\rm PCTr} \: \subseteq \: \sbis{\rm PCTr,\sqcup\sqcap}$.

\item $\sbis{\rm PF,dis} \: \subseteq \: \sbis{\rm PCTr,dis} \: \subseteq \: \sbis{\rm PTr,dis}$.

\item $\sbis{\rm PF} \: \subseteq \: \sbis{\rm PCTr} \: \subseteq \: \sbis{\rm PTr}$.

\item $\sbis{\rm PF,\sqcup\sqcap} \: \subseteq \: \sbis{\rm PCTr,\sqcup\sqcap} \: \subseteq \: \sbis{\rm
PTr,\sqcup\sqcap}$.

		\end{enumerate}

		\begin{proof}
Let $(S, A, \! \arrow{}{} \!)$ be an NPLTS and $s_{1}, s_{2} \in S$:

			\begin{enumerate}

\item Similar to the proof of Thm.~3.5 in~\cite{BDL13TCS}.

\item Suppose that $s_{1} \sbis{\rm PF,dis} s_{2}$. Then we immediately derive that:

				\begin{itemize}

\item For each $\calz_{1} \in \ms{Res}(s_{1})$ there exists $\calz_{2} \in \ms{Res}(s_{2})$ such that for
all $\alpha \in A^{*}$:
\cws{0}{\hspace*{-1.6cm}\begin{array}{rcccl}
\ms{prob}(\calcc(z_{s_{1}}, \alpha)) & \!\!\! = \!\!\! & \ms{prob}(\calfcc(z_{s_{1}}, (\alpha, \emptyset)))
& \!\!\! = \!\!\! & \\
& \!\!\! = \!\!\! & \ms{prob}(\calfcc(z_{s_{2}}, (\alpha, \emptyset))) & \!\!\! = \!\!\! &
\ms{prob}(\calcc(z_{s_{2}}, \alpha)) \\
\ms{prob}(\calccc(z_{s_{1}}, \alpha)) & \!\!\! = \!\!\! & \ms{prob}(\calfcc(z_{s_{1}}, (\alpha, A))) &
\!\!\! = \!\!\! & \\
& \!\!\! = \!\!\! & \ms{prob}(\calfcc(z_{s_{2}}, (\alpha, A))) & \!\!\! = \!\!\! &
\ms{prob}(\calccc(z_{s_{2}}, \alpha)) \\
\end{array}}

\item Symmetrically for each $\calz_{2} \in \ms{Res}(s_{2})$.

				\end{itemize}

This means that $s_{1} \sbis{\rm PCTr,dis} s_{2}$. \\
The fact that $s_{1} \sbis{\rm PCTr,dis} s_{2}$ implies $s_{1} \sbis{\rm PTr,dis} s_{2}$ is a
straightforward consequence of the definition of the two equivalences.

\item Suppose that $s_{1} \sbis{\rm PF} s_{2}$. Then we immediately derive that for all $\alpha \in A^{*}$:

				\begin{itemize}

\item For each $\calz_{1} \in \ms{Res}(s_{1})$ there exist $\calz_{2} \in \ms{Res}(s_{2})$ such that:
\cws{0}{\hspace*{-1.6cm}\begin{array}{rcccl}
\ms{prob}(\calcc(z_{s_{1}}, \alpha)) & \!\!\! = \!\!\! & \ms{prob}(\calfcc(z_{s_{1}}, (\alpha, \emptyset)))
& \!\!\! = \!\!\! & \\
& \!\!\! = \!\!\! & \ms{prob}(\calfcc(z_{s_{2}}, (\alpha, \emptyset))) & \!\!\! = \!\!\! &
\ms{prob}(\calcc(z_{s_{2}}, \alpha)) \\
\end{array}}
and $\calz'_{2} \in \ms{Res}(s_{2})$ such that:
\cws{0}{\hspace*{-1.6cm}\begin{array}{rcccl}
\ms{prob}(\calccc(z_{s_{1}}, \alpha)) & \!\!\! = \!\!\! & \ms{prob}(\calfcc(z_{s_{1}}, (\alpha, A))) &
\!\!\! = \!\!\! & \\
& \!\!\! = \!\!\! & \ms{prob}(\calfcc(z'_{s_{2}}, (\alpha, A))) & \!\!\! = \!\!\! &
\ms{prob}(\calccc(z'_{s_{2}}, \alpha)) \\
\end{array}}

\item Symmetrically for each $\calz_{2} \in \ms{Res}(s_{2})$.

				\end{itemize}

This means that $s_{1} \sbis{\rm PCTr} s_{2}$. \\
The fact that $s_{1} \sbis{\rm PCTr} s_{2}$ implies $s_{1} \sbis{\rm PTr} s_{2}$ is a straightforward
consequence of the definition of the two equivalences.

\item Suppose that $s_{1} \sbis{\rm PF,\sqcup\sqcap} s_{2}$. Then we immediately derive that for all $\alpha
\in A^{*}$:
\cws{0}{\hspace*{-0.8cm}\begin{array}{rcl}
\bigsqcup\limits_{\calz_{1} \in \ms{Res}_{\alpha}(s_{1})} \ms{prob}(\calcc(z_{s_{1}}, \alpha)) & \!\!\! =
\!\!\! & \bigsqcup\limits_{\calz_{1} \in \ms{Res}_{\alpha}(s_{1})} \ms{prob}(\calfcc(z_{s_{1}}, (\alpha,
\emptyset))) \\[0.4cm]
& \!\!\! = \!\!\! & \bigsqcup\limits_{\calz_{2} \in \ms{Res}_{\alpha}(s_{2})} \ms{prob}(\calfcc(z_{s_{2}},
(\alpha, \emptyset))) \\[0.4cm]
& \!\!\! = \!\!\! & \bigsqcup\limits_{\calz_{2} \in \ms{Res}_{\alpha}(s_{2})} \ms{prob}(\calcc(z_{s_{2}},
\alpha)) \\[0.4cm]
\bigsqcap\limits_{\calz_{1} \in \ms{Res}_{\alpha}(s_{1})} \ms{prob}(\calcc(z_{s_{1}}, \alpha)) & \!\!\! =
\!\!\! & \bigsqcap\limits_{\calz_{1} \in \ms{Res}_{\alpha}(s_{1})} \ms{prob}(\calfcc(z_{s_{1}}, (\alpha,
\emptyset))) \\[0.4cm]
& \!\!\! = \!\!\! & \bigsqcap\limits_{\calz_{2} \in \ms{Res}_{\alpha}(s_{2})} \ms{prob}(\calfcc(z_{s_{2}},
(\alpha, \emptyset))) \\[0.4cm]
& \!\!\! = \!\!\! & \bigsqcap\limits_{\calz_{2} \in \ms{Res}_{\alpha}(s_{2})} \ms{prob}(\calcc(z_{s_{2}},
\alpha)) \\
\end{array}}
and:
\cws{0}{\hspace*{-0.8cm}\begin{array}{rcl}
\bigsqcup\limits_{\calz_{1} \in \ms{Res}_{\alpha}(s_{1})} \ms{prob}(\calccc(z_{s_{1}}, \alpha)) & \!\!\! =
\!\!\! & \bigsqcup\limits_{\calz_{1} \in \ms{Res}_{\alpha}(s_{1})} \ms{prob}(\calfcc(z_{s_{1}}, (\alpha,
A))) \\[0.4cm]
& \!\!\! = \!\!\! & \bigsqcup\limits_{\calz_{2} \in \ms{Res}_{\alpha}(s_{2})} \ms{prob}(\calfcc(z_{s_{2}},
(\alpha, A))) \\[0.4cm]
& \!\!\! = \!\!\! & \bigsqcup\limits_{\calz_{2} \in \ms{Res}_{\alpha}(s_{2})} \ms{prob}(\calccc(z_{s_{2}},
\alpha)) \\[0.4cm]
\bigsqcap\limits_{\calz_{1} \in \ms{Res}_{\alpha}(s_{1})} \ms{prob}(\calccc(z_{s_{1}}, \alpha)) & \!\!\! =
\!\!\! & \bigsqcap\limits_{\calz_{1} \in \ms{Res}_{\alpha}(s_{1})} \ms{prob}(\calfcc(z_{s_{1}}, (\alpha,
A))) \\[0.4cm]
& \!\!\! = \!\!\! & \bigsqcap\limits_{\calz_{2} \in \ms{Res}_{\alpha}(s_{2})} \ms{prob}(\calfcc(z_{s_{2}},
(\alpha, A))) \\[0.4cm]
& \!\!\! = \!\!\! & \bigsqcap\limits_{\calz_{2} \in \ms{Res}_{\alpha}(s_{2})} \ms{prob}(\calccc(z_{s_{2}},
\alpha)) \\
\end{array}}
This means that $s_{1} \sbis{\rm PCTr,\sqcup\sqcap} s_{2}$. \\
The fact that $s_{1} \sbis{\rm PCTr,\sqcup\sqcap} s_{2}$ implies $s_{1} \sbis{\rm PTr,\sqcup\sqcap} s_{2}$
is a straightforward consequence of the definition of the two equivalences.
\fullbox

			\end{enumerate}

		\end{proof}

	\end{theorem}

	\begin{figure}[tp]

\input{Pictures/dis_vs_by}
\caption{Two NPLTS models distinguished by equivalences in fragment 1 and identified by those in fragment 2}
\label{fig:dis_vs_by}

	\end{figure}

	\begin{figure}[tp]

\input{Pictures/by_vs_supinf}
\caption{Two NPLTS models distinguished by equivalences in fragment 2 and identified by those in fragment 3}
\label{fig:by_vs_supinf}

	\end{figure}
	
All the inclusions in Thm.~\ref{thm:compl_trace_results} are strict:

	\begin{itemize}

\item Figures~\ref{fig:dis_vs_by} and~\ref{fig:by_vs_supinf} respectively show that $\sbis{\rm PCTr,dis}$ is
strictly finer than $\sbis{\rm PCTr}$ and $\sbis{\rm PCTr}$ is strictly finer than $\sbis{\rm
PCTr,\sqcup\sqcap}$.

	\begin{figure}[tp]

\input{Pictures/pf_vs_ptr}
\caption{Two NPLTS models distinguished by $\sbis{\rm PF,dis}$/$\sbis{\rm PF}$/$\sbis{\rm PF,\sqcup\sqcap}$
and identified by $\sbis{\rm PCTr,dis}$/$\sbis{\rm PCTr}$/$\sbis{\rm PCTr,\sqcup\sqcap}$}
\label{fig:pf_vs_ptr}

	\end{figure}

	\begin{figure}[tp]

\input{Pictures/ptrsupinf_anomaly_maxres}
\caption{Two NPLTS models distinguished by $\sbis{\rm PCTr,dis}$/$\sbis{\rm PCTr}$/$\sbis{\rm
PCTr,\sqcup\sqcap}$ and identified by $\sbis{\rm PTr,dis}$/$\sbis{\rm PTr}$/$\sbis{\rm PTr,\sqcup\sqcap}$}
\label{fig:ptrsupinf_anomaly_maxres}

	\end{figure}

\item Figure~\ref{fig:pf_vs_ptr} shows that $\sbis{\rm PF,dis}$, $\sbis{\rm PF}$, and $\sbis{\rm
PF,\sqcup\sqcap}$ are strictly finer than $\sbis{\rm PCTr,dis}$, $\sbis{\rm PCTr}$, and $\sbis{\rm
PCTr,\sqcup\sqcap}$, respectively. Indeed, for each resolution of~$s_{1}$ (resp.\ $s_{2}$) there exists a
resolution of $s_{2}$ (resp.\ $s_{1}$) such that both resolutions have precisely the same trace distribution
and the same completed-trace distribution, thus $s_{1}$ and $s_{2}$ are identified by $\sbis{\rm PCTr,dis}$
(and hence by $\sbis{\rm PCTr}$ and $\sbis{\rm PCTr,\sqcup\sqcap}$). In contrast, the leftmost
$a$-computation of $s_{1}$ is compatible with the failure pair $(a, \{ c \})$ while $s_{2}$ has no
computation compatible with that failure pair, thus $s_{1}$ and $s_{2}$ are distinguished by $\sbis{\rm
PF,\sqcup\sqcap}$ (and hence by $\sbis{\rm PF}$ and $\sbis{\rm PF,dis}$).

\item Figure~\ref{fig:ptrsupinf_anomaly_maxres} shows that $\sbis{\rm PCTr,dis}$, $\sbis{\rm PCTr}$, and
$\sbis{\rm PCTr,\sqcup\sqcap}$ are strictly finer than $\sbis{\rm PTr,dis}$, $\sbis{\rm PTr}$, and
$\sbis{\rm PTr,\sqcup\sqcap}$, respectively. Indeed, for each resolution of~$s_{1}$ (resp.\ $s_{2}$) there
exists a resolution of $s_{2}$ (resp.\ $s_{1}$) such that both resolutions have precisely the same trace
distribution, thus $s_{1}$ and $s_{2}$ are identified by $\sbis{\rm PTr,dis}$ (and hence by $\sbis{\rm PTr}$
and $\sbis{\rm PTr,\sqcup\sqcap}$). In contrast, the rightmost $a$-computation of $s_{1}$ is completed while
$s_{2}$ has no completed $a$-compatible computation, thus $s_{1}$ and $s_{2}$ are distinguished by
$\sbis{\rm PCTr,\sqcup\sqcap}$ (and hence by $\sbis{\rm PCTr}$ and $\sbis{\rm PCTr,dis}$).

	\end{itemize}

Moreover:

	\begin{itemize}

\item $\sbis{\rm PB}$ and $\sbis{\rm PB,\sqcup\sqcap}$ are incomparable with the three completed-trace
equivalences. Indeed, in Fig.~\ref{fig:pb_pbsupinf_vs_others} it holds that $s_{1} \sbis{\rm PB} s_{2}$ (and
hence $s_{1} \sbis{\rm PB,\sqcup\sqcap} s_{2}$) -- as can be seen by taking the equivalence relation that
pairs states having equally labeled transitions leading to the same distribution -- and $s_{1} \not\sbis{\rm
PCTr,\sqcup\sqcap} s_{2}$ (and hence $s_{1} \not\sbis{\rm PCTr} s_{2}$ and $s_{1} \not\sbis{\rm PCTr,dis}
s_{2}$) -- due to the trace $a \, b \, c$ having maximum probability $0.68$ in the first process and $0.61$
in the second process. In contrast, in Fig.~\ref{fig:pbdis_vs_ptetbtdis} it holds that $s_{1} \not\sbis{\rm
PB,\sqcup\sqcap} s_{2}$ (and hence $s_{1} \not\sbis{\rm PB} s_{2}$) -- as the leftmost state with outgoing
$b$-transitions reachable from $s_{2}$ is not $\sqcup\sqcap$-bisimilar to the two states with outgoing
$b$-transitions reachable from $s_{1}$ -- and $s_{1} \sbis{\rm PCTr,dis} s_{2}$ (and hence $s_{1}
\sbis{\rm PCTr} s_{2}$ and $s_{1} \sbis{\rm PCTr,\sqcup\sqcap} s_{2}$).

	\begin{figure}[tp]

\input{Pictures/pb_pbsupinf_vs_others}
\caption{Two NPLTS models distinguished by $\sbis{\rm PCTr,dis}$/$\sbis{\rm PCTr}$/$\sbis{\rm
PCTr,\sqcup\sqcap}$ and identified by $\sbis{\rm PB}$/$\sbis{\rm PB,\sqcup\sqcap}$}
\label{fig:pb_pbsupinf_vs_others}

	\end{figure}

	\begin{figure}[tp]

\input{Pictures/pbdis_vs_ptetbtdis}
\caption{Two NPLTS models distinguished by $\sbis{\rm PB}$/$\sbis{\rm PB,\sqcup\sqcap}$ and identified by
$\sbis{\rm PCTr,dis}$/$\sbis{\rm PCTr}$/$\sbis{\rm PCTr,\sqcup\sqcap}$}
\label{fig:pbdis_vs_ptetbtdis}

	\end{figure}

	\begin{figure}[bp]

\input{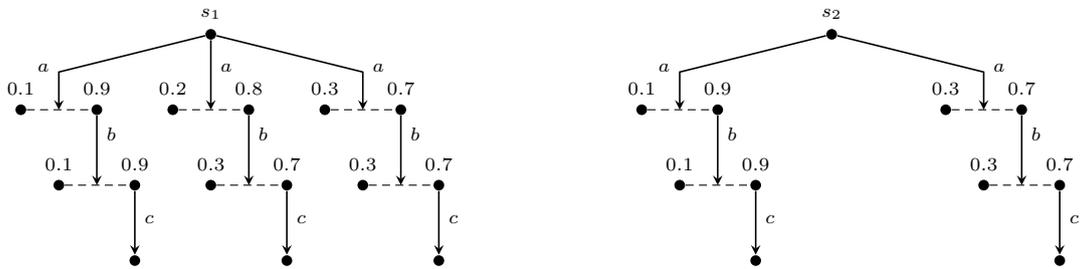}
\caption{Two NPLTS models distinguished by $\sbis{\rm PCTr,\sqcup\sqcap}$ and identified by
$\sbis{\textrm{PTe-}\sqcup\sqcap}$}
\label{fig:pfsupinf_vs_ptesupinf}

	\end{figure}

\item $\sbis{\textrm{PTe-}\sqcup\sqcap}$ is incomparable with the three completed-trace equivalences.
Indeed, in Fig.~\ref{fig:by_vs_supinf} it holds that $s_{1} \sbis{\textrm{PTe-}\sqcup\sqcap} s_{2}$ and
$s_{1} \not\sbis{\rm PCTr} s_{2}$ (and hence $s_{1} \not\sbis{\rm PCTr,dis}$). In contrast, in
Fig.~\ref{fig:ptesupinf_anomaly} it holds that $s_{1} \not\sbis{\textrm{PTe-}\sqcup\sqcap} s_{2}$ -- due to
the test shown in the figure -- and $s_{1} \sbis{\rm PCTr,dis} s_{2}$ (and hence $s_{1} \sbis{\rm PCTr}
s_{2}$). Likewise, in Fig.~\ref{fig:pfsupinf_vs_ptesupinf} it holds that $s_{1}
\sbis{\textrm{PTe-}\sqcup\sqcap} s_{2}$ -- as there is no test that results in an interaction system having
a maximal resolution with differently labeled successful computations of the same length and hence no
possibility of summing up their success probabilities -- and $s_{1} \not\sbis{\rm PCTr,\sqcup\sqcap} s_{2}$
-- due to the completed trace $a \, b$ whose maximum probability is $0.24$ in the first process and $0.21$
in the second process. In contrast, in Fig.~\ref{fig:dis_vs_by} it holds that $s_{1}
\not\sbis{\textrm{PTe-}\sqcup\sqcap} s_{2}$ and $s_{1} \sbis{\rm PCTr,\sqcup\sqcap} s_{2}$.

	\begin{figure}[tp]

\input{Pictures/ptesupinf_anomaly}
\caption{Two NPLTS models distinguished by $\sbis{\textrm{PTe-}\sqcup\sqcap}$ and identified by $\sbis{\rm
PCTr,dis}$/$\sbis{\rm PCTr}$}
\label{fig:ptesupinf_anomaly}

	\end{figure}

	\begin{figure}[tp]

\input{Pictures/pfsupinf_vs_ptetbtsupinf}
\caption{Two NPLTS models distinguished by $\sbis{\rm PCTr,dis}$/$\sbis{\rm PCTr}$/$\sbis{\rm
PCTr,\sqcup\sqcap}$ and identified by $\sbis{\textrm{PTe-tbt}}$/$\sbis{\textrm{PTe-tbt},\sqcup\sqcap}$}
\label{fig:pfsupinf_vs_ptetbtsupinf}

	\end{figure}

	\begin{figure}[tp]

\input{Pictures/ptetbtsupinf_vs_ptrsupinf}
\caption{Two NPLTS models distinguished by
$\sbis{\textrm{PTe-tbt}}$/$\sbis{\textrm{PTe-tbt,$\sqcup\sqcap$}}$ and identified by $\sbis{\rm
PCTr,dis}$/$\sbis{\rm PCTr}$/$\sbis{\rm PCTr,\sqcup\sqcap}$}
\label{fig:ptetbtsupinf_vs_ptrsupinf}

	\end{figure}

\item $\sbis{\textrm{PTe-tbt}}$ and $\sbis{\textrm{PTe-tbt,$\sqcup\sqcap$}}$ are incomparable with the three
completed-trace equivalences, because in Fig.~\ref{fig:pfsupinf_vs_ptetbtsupinf} it holds that $s_{1}
\sbis{\textrm{PTe-tbt}} s_{2}$ (and hence $s_{1} \sbis{\textrm{PTe-tbt,$\sqcup\sqcap$}} s_{2}$) and $s_{1}
\not\sbis{\rm PCTr,\sqcup\sqcap} s_{2}$ (and hence $s_{1} \not\sbis{\rm PCTr} s_{2}$ and $s_{1}
\not\sbis{\rm PCTr,dis} s_{2}$), while in Fig.~\ref{fig:ptetbtsupinf_vs_ptrsupinf} it holds that $s_{1}
\not\sbis{\textrm{PTe-tbt,$\sqcup\sqcap$}} s_{2}$ (and hence \linebreak $s_{1} \not\sbis{\textrm{PTe-tbt}}
s_{2}$) and $s_{1} \sbis{\rm PCTr,dis} s_{2}$ (and hence $s_{1} \sbis{\rm PCTr} s_{2}$ and $s_{1} \sbis{\rm
PCTr,\sqcup\sqcap} s_{2}$).

\item $\sbis{\rm PF}$ and $\sbis{\rm PF,\sqcup\sqcap}$ are incomparable with $\sbis{\rm PCTr,dis}$, because
in Fig.~\ref{fig:dis_vs_by} it holds that $s_{1} \sbis{\rm PF} s_{2}$ (and hence $s_{1} \sbis{\rm
PF,\sqcup\sqcap} s_{2}$) and $s_{1} \not\sbis{\rm PCTr,dis} s_{2}$, while in Fig.~\ref{fig:pf_vs_ptr} it
holds that $s_{1} \not\sbis{\rm PF,\sqcup\sqcap} s_{2}$ (and hence $s_{1} \not\sbis{\rm PF} s_{2}$) and
$s_{1} \sbis{\rm PCTr,dis} s_{2}$.

\item $\sbis{\rm PF,\sqcup\sqcap}$ is incomparable with $\sbis{\rm PCTr}$, because in
Fig.~\ref{fig:by_vs_supinf} it holds that $s_{1} \sbis{\rm PF,\sqcup\sqcap} s_{2}$ and $s_{1} \not\sbis{\rm
PCTr} s_{2}$, while in Fig.~\ref{fig:pf_vs_ptr} it holds that $s_{1} \not\sbis{\rm PF,\sqcup\sqcap} s_{2}$
and $s_{1} \sbis{\rm PCTr} s_{2}$.

\item $\sbis{\rm PCTr}$ and $\sbis{\rm PCTr,\sqcup\sqcap}$ are incomparable with $\sbis{\rm PTr,dis}$,
because in Fig.~\ref{fig:dis_vs_by} it holds that $s_{1} \sbis{\rm PCTr} s_{2}$ (and hence $s_{1} \sbis{\rm
PCTr,\sqcup\sqcap} s_{2}$) and $s_{1} \not\sbis{\rm PTr,dis} s_{2}$, while in
Fig.~\ref{fig:ptrsupinf_anomaly_maxres} it holds that $s_{1} \not\sbis{\rm PCTr,\sqcup\sqcap} s_{2}$ (and
hence $s_{1} \not\sbis{\rm PCTr} s_{2}$) and $s_{1} \sbis{\rm PTr,dis} s_{2}$.

\item $\sbis{\rm PCTr,\sqcup\sqcap}$ is incomparable with $\sbis{\rm PTr}$, because in
Fig.~\ref{fig:by_vs_supinf} it holds that $s_{1} \sbis{\rm PCTr,\sqcup\sqcap} s_{2}$ and $s_{1}
\not\sbis{\rm PTr} s_{2}$, while in Fig.~\ref{fig:ptrsupinf_anomaly_maxres} it holds that $s_{1}
\not\sbis{\rm PCTr,\sqcup\sqcap} s_{2}$ and $s_{1} \sbis{\rm PTr} s_{2}$.

	\end{itemize}

%
\subsection{Failure-Trace, Readiness, and Ready-Trace Equivalences}\label{sec:decor_trace_equiv}
%

\noindent
Failure semantics generalizes completed-trace equivalence towards arbitrary safety properties. An extension
of failure semantics is failure-trace semantics. We call \emph{failure trace} an element $\phi \in (A \times
2^{A})^{*}$ given by a sequence of $n \in \natns$ pairs of the form $(a_{i}, F_{i})$. We say that $c \in
\calc_{\rm fin}(z_{s})$ is \emph{compatible} with $\phi$ iff $c \in \calcc(z_{s}, a_{1} \dots a_{n})$ and,
denoting by $z_{i}$ the state reached by~$c$ after the $i$-th step for all $i = 1, \dots, n$,
$\ms{corr}_{\calz}(z_{i})$ has no outgoing transitions in $\call$ labeled with an action in $F_{i}$. We
denote by $\calftcc(z_{s}, \phi)$ the set of \linebreak $\phi$-compatible computations from $z_{s}$.

	\begin{definition}\label{def:pftrdis}

(\emph{Probabilistic failure-trace-distribution equivalence} -- $\sbis{\rm PFTr,dis}$) \\
$s_{1} \sbis{\rm PFTr,dis} s_{2}$ iff for each $\calz_{1} \in \ms{Res}(s_{1})$ there exists $\calz_{2} \in
\ms{Res}(s_{2})$ such that \underline{for all $\phi \in (A \times 2^{A})^{*}$}:
\cws{0}{\ms{prob}(\calftcc(z_{s_{1}}, \phi)) \: = \: \ms{prob}(\calftcc(z_{s_{2}}, \phi))}
and symmetrically for each $\calz_{2} \in \ms{Res}(s_{2})$.
\fullbox

	\end{definition}

	\begin{definition}\label{def:pftr}

(\emph{Probabilistic failure-trace equivalence} -- $\sbis{\rm PFTr}$) \\
$s_{1} \sbis{\rm PFTr} s_{2}$ iff \underline{for all $\phi \in (A \times 2^{A})^{*}$} it holds that for each
$\calz_{1} \in \ms{Res}(s_{1})$ there exists $\calz_{2} \in \ms{Res}(s_{2})$ such that:
\cws{0}{\ms{prob}(\calftcc(z_{s_{1}}, \phi)) \: = \: \ms{prob}(\calftcc(z_{s_{2}}, \phi))}
and symmetrically for each $\calz_{2} \in \ms{Res}(s_{2})$.
\fullbox

	\end{definition}

	\begin{definition}\label{def:pftrsupinf}

(\emph{Probabilistic $\sqcup\sqcap$-failure-trace equivalence} -- $\sbis{\rm PFTr,\sqcup\sqcap}$) \\
$s_{1} \sbis{\rm PF,\sqcup\sqcap} s_{2}$ iff for all $\phi \in (A \times 2^{A})^{*}$:
\cws{10}{\begin{array}{rcl}
\bigsqcup\limits_{\calz_{1} \in \ms{Res}_{\alpha}(s_{1})} \ms{prob}(\calftcc(z_{s_{1}}, \phi)) & \!\!\! =
\!\!\! & \bigsqcup\limits_{\calz_{2} \in \ms{Res}_{\alpha}(s_{2})} \ms{prob}(\calftcc(z_{s_{2}}, \phi))
\\[0.4cm]
\bigsqcap\limits_{\calz_{1} \in \ms{Res}_{\alpha}(s_{1})} \ms{prob}(\calftcc(z_{s_{1}}, \phi)) & \!\!\! =
\!\!\! & \bigsqcap\limits_{\calz_{2} \in \ms{Res}_{\alpha}(s_{2})} \ms{prob}(\calftcc(z_{s_{2}}, \phi)) \\
\end{array}}
\fullbox

	\end{definition}

A different generalization towards liveness properties is readiness semantics, which considers the set of
actions that can be accepted after performing a trace. We call \emph{ready pair} an element $\varrho \in
A^{*} \times 2^{A}$ formed by a trace $\alpha$ and a decoration~$R$ called \emph{ready set}. We say that $c$
is \emph{compatible} with~$\varrho$ iff $c \in \calcc(z_{s}, \alpha)$ and the set of actions labeling the
transitions in $\call$ departing from $\ms{corr}_{\calz}(\ms{last}(c))$ is precisely~$R$. We denote by
$\calrcc(z_{s}, \varrho)$ the set of $\varrho$-compatible computations from $z_{s}$.

	\begin{definition}\label{def:prdis}

(\emph{Probabilistic readiness-distribution equivalence} -- $\sbis{\rm PR,dis}$) \\
$s_{1} \sbis{\rm PR,dis} s_{2}$ iff for each $\calz_{1} \in \ms{Res}(s_{1})$ there exists $\calz_{2} \in
\ms{Res}(s_{2})$ such that \underline{for all $\varrho \in A^{*} \times 2^{A}$}:
\cws{0}{\ms{prob}(\calrcc(z_{s_{1}}, \varrho)) \: = \: \ms{prob}(\calrcc(z_{s_{2}}, \varrho))}
and symmetrically for each $\calz_{2} \in \ms{Res}(s_{2})$.
\fullbox

	\end{definition}

	\begin{definition}\label{def:pr}

(\emph{Probabilistic readiness equivalence} -- $\sbis{\rm PR}$) \\
$s_{1} \sbis{\rm PR} s_{2}$ iff \underline{for all $\varrho \in A^{*} \times 2^{A}$} it holds that for each
$\calz_{1} \in \ms{Res}(s_{1})$ there exists $\calz_{2} \in \ms{Res}(s_{2})$ such that:
\cws{0}{\ms{prob}(\calrcc(z_{s_{1}}, \varrho)) \: = \: \ms{prob}(\calrcc(z_{s_{2}}, \varrho))}
and symmetrically for each $\calz_{2} \in \ms{Res}(s_{2})$.
\fullbox

	\end{definition}

	\begin{definition}\label{def:prsupinf}

(\emph{Probabilistic $\sqcup\sqcap$-readiness equivalence} -- $\sbis{\rm PR,\sqcup\sqcap}$) \\
$s_{1} \sbis{\rm PR,\sqcup\sqcap} s_{2}$ iff for all $\varrho = (\alpha, R) \in A^{*} \times 2^{A}$:
\cws{10}{\begin{array}{rcl}
\bigsqcup\limits_{\calz_{1} \in \ms{Res}_{\alpha}(s_{1})} \ms{prob}(\calrcc(z_{s_{1}}, \varrho)) & \!\!\! =
\!\!\! & \bigsqcup\limits_{\calz_{2} \in \ms{Res}_{\alpha}(s_{2})} \ms{prob}(\calrcc(z_{s_{2}}, \varrho))
\\[0.4cm]
\bigsqcap\limits_{\calz_{1} \in \ms{Res}_{\alpha}(s_{1})} \ms{prob}(\calrcc(z_{s_{1}}, \varrho)) & \!\!\! =
\!\!\! & \bigsqcap\limits_{\calz_{2} \in \ms{Res}_{\alpha}(s_{2})} \ms{prob}(\calrcc(z_{s_{2}}, \varrho)) \\
\end{array}}
\fullbox

	\end{definition}

Moreover, we call \emph{ready trace} an element $\rho \in (A \times 2^{A})^{*}$ given by a sequence of $n
\in \natns$ pairs of the form $(a_{i}, R_{i})$. We say that $c \in \calc_{\rm fin}(z_{s})$ is
\emph{compatible} with $\rho$ iff $c \in \calcc(z_{s}, a_{1} \dots a_{n})$ and, denoting by $z_{i}$ the
state reached by~$c$ after the $i$-th step for all $i = 1, \dots, n$, the set of actions labeling the
transitions in $\call$ departing from $\ms{corr}_{\calz}(z_{i})$ is precisely $R_{i}$. We denote by
$\calrtcc(z_{s}, \rho)$ the set of $\rho$-compatible computations from $z_{s}$.

	\begin{definition}\label{def:prtrdis}

(\emph{Probabilistic ready-trace-distribution equivalence} -- $\sbis{\rm PRTr,dis}$) \\
$s_{1} \sbis{\rm PRTr,dis} s_{2}$ iff for each $\calz_{1} \in \ms{Res}(s_{1})$ there exists $\calz_{2} \in
\ms{Res}(s_{2})$ such that \underline{for all $\rho \in (A \times 2^{A})^{*}$}:
\cws{0}{\ms{prob}(\calrtcc(z_{s_{1}}, \rho)) \: = \: \ms{prob}(\calrtcc(z_{s_{2}}, \rho))}
and symmetrically for each $\calz_{2} \in \ms{Res}(s_{2})$.
\fullbox

	\end{definition}

	\begin{definition}\label{def:prtr}

(\emph{Probabilistic ready-trace equivalence} -- $\sbis{\rm PRTr}$) \\
$s_{1} \sbis{\rm PRTr} s_{2}$ iff \underline{for all $\rho \in (A \times 2^{A})^{*}$} it holds that for each
$\calz_{1} \in \ms{Res}(s_{1})$ there exists $\calz_{2} \in \ms{Res}(s_{2})$ such that:
\cws{0}{\ms{prob}(\calrtcc(z_{s_{1}}, \rho)) \: = \: \ms{prob}(\calrtcc(z_{s_{2}}, \rho))}
and symmetrically for each $\calz_{2} \in \ms{Res}(s_{2})$.
\fullbox

	\end{definition}

	\begin{definition}\label{def:prtrsupinf}

(\emph{Probabilistic $\sqcup\sqcap$-ready-trace equivalence} -- $\sbis{\rm PRTr,\sqcup\sqcap}$) \\
$s_{1} \sbis{\rm PRTr,\sqcup\sqcap} s_{2}$ iff for all $\rho \in (A \times 2^{A})^{*}$:
\cws{10}{\begin{array}{rcl}
\bigsqcup\limits_{\calz_{1} \in \ms{Res}_{\alpha}(s_{1})} \ms{prob}(\calrtcc(z_{s_{1}}, \rho)) & \!\!\! =
\!\!\! & \bigsqcup\limits_{\calz_{2} \in \ms{Res}_{\alpha}(s_{2})} \ms{prob}(\calrtcc(z_{s_{2}}, \rho))
\\[0.4cm]
\bigsqcap\limits_{\calz_{1} \in \ms{Res}_{\alpha}(s_{1})} \ms{prob}(\calrtcc(z_{s_{1}}, \rho)) & \!\!\! =
\!\!\! & \bigsqcap\limits_{\calz_{2} \in \ms{Res}_{\alpha}(s_{2})} \ms{prob}(\calrtcc(z_{s_{2}}, \rho)) \\
\end{array}}
\fullbox

	\end{definition}

We now investigate the relationships of the nine additional decorated-trace equivalences among themselves
and with the various equivalences defined in~\cite{BDL13TCS} and in this paper. As in the fully
probabilistic spectrum~\cite{JS90,HT92}, for the decorated-trace equivalences based on fully matching
resolutions it holds that readiness semantics coincides with failure semantics, and this extends to
ready-trace semantics and failure-trace semantics. In contrast, for the other decorated-trace equivalences
based on partially matching resolutions or extremal probabilities, unlike the fully nondeterministic
spectrum~\cite{Gla01} it turns out that ready-trace semantics and readiness semantic are incomparable with
most of the other semantics.

	\begin{theorem}\label{thm:decor_trace_results}

It holds that:

		\begin{enumerate}

\item $\sbis{\rm \pi,dis} \: \subseteq \: \sbis{\rm \pi} \: \subseteq \: \sbis{\rm \pi,\sqcup\sqcap}$ for
all $\pi \in \{ {\rm PRTr}, {\rm PFTr}, {\rm PR} \}$.

\item $\sbis{\textrm{PTe-tbt,dis}} \: \subseteq \: \sbis{\rm PRTr,dis}$.

\item $\sbis{\rm PRTr,dis} \: = \: \sbis{\rm PFTr,dis}$ over finitely-branching NPLTS models.

\item $\sbis{\rm PR,dis} \: = \: \sbis{\rm PF,dis}$ over finitely-branching NPLTS models.

\item $\sbis{\rm PFTr,dis} \: \subseteq \: \sbis{\rm PF,dis}$.

\item $\sbis{\rm PFTr} \: \subseteq \: \sbis{\rm PF}$.

\item $\sbis{\rm PFTr,\sqcup\sqcap} \: \subseteq \: \sbis{\rm PF,\sqcup\sqcap}$.

		\end{enumerate}

		\begin{proof}
Let $(S, A, \! \arrow{}{} \!)$ be an NPLTS and $s_{1}, s_{2} \in S$:

			\begin{enumerate}

\item Similar to the proof of Thm.~3.5 in~\cite{BDL13TCS}.

\item We show that $s_{1} \sbis{\textrm{PTe-tbt,dis}} s_{2}$ implies $s_{1} \sbis{\rm PRTr,dis} s_{2}$ by
building a test that permits to reason about all ready traces at once for each resolution of $s_{1}$ and
$s_{2}$. We start by deriving a new NPLTS $(S_{\rm r}, A_{\rm r}, \! \arrow{}{\rm r} \!)$ that is isomorphic
to the given one up to transition labels and terminal states. A transition $s \arrow{a}{} \cald$ becomes
$s_{\rm r} \arrow{a \triangleleft R}{\rm r} \cald_{\rm r}$ where $R \subseteq A$ is the set of actions
labeling the outgoing transitions of~$s$ and $\cald_{\rm r}(s_{\rm r}) = \cald(s)$ for all $s \in S$. If $s$
is a terminal state, i.e., it has no outgoing transitions, then we add a transition $s_{\rm r} \arrow{\circ
\triangleleft \emptyset}{\rm r} \delta_{s_{\rm r}}$ where $\delta_{s_{\rm r}}(s_{\rm r}) = 1$ and
$\delta_{s_{\rm r}}(s'_{\rm r}) = 0$ for all $s' \in S \setminus \{ s \}$. Transition relabeling preserves
$\sbis{\textrm{PTe-tbt,dis}}$, i.e., $s_{1} \sbis{\textrm{PTe-tbt,dis}} s_{2}$ implies $s_{1, \rm r}
\sbis{\textrm{PTe-tbt,dis}} s_{2, \rm r}$, because $\sbis{\textrm{PTe-tbt,dis}}$ is able to distinguish a
state that has a single $\alpha$-compatible computation reaching a state with a nondeterministic branching
formed by a $b$-transition and a $c$-transition, from a state that has two $\alpha$-compatible computations
such that one of them reaches a state with only one outgoing transition labeled with $b$ and the other one
reaches a state with only one outgoing transition labeled with $c$ (e.g., use a test that has a single
$\alpha$-compatible computation whose last step leads to a distribution whose support contains only a state
with only one outgoing transition labeled with $b$ that reaches success and a state with only one outgoing
transition labeled with~$c$ that reaches success). \\
For each $\alpha_{\rm r} \in (A_{\rm r})^{*}$ and $R \subseteq A$, we build an NPT $\calt_{\alpha_{\rm r},
R} = (O_{\alpha_{\rm r}, R}, A_{\rm r}, \! \arrow{}{\alpha_{\rm r}, R} \!)$ having a single \linebreak
$\alpha_{\rm r}$-compatible computation that goes from the initial state $o_{\alpha_{\rm r}, R}$ to a state
having a single transition to $\omega$ labeled with (i) $\circ \triangleleft \emptyset$ if $R = \emptyset$
or (ii) $\_ \triangleleft R$ if $R \neq \emptyset$. Since we compare individual states (like $s_{1}$
and~$s_{2}$) rather than state distributions, the distinguishing power of $\sbis{\textrm{PTe-tbt,dis}}$ does
not change if we additionally consider tests starting with a single $\tau$-transition that can initially
evolve autonomously in any interaction system. We thus build a further NPT $\calt = (O, A_{\rm r}, \!
\arrow{}{\calt} \!)$ that has an initial $\tau$-transition and then behaves as one of the tests
$\calt_{\alpha_{\rm r}, R}$, i.e., its initial $\tau$-transition goes from the initial state $o$ to a state
distribution whose support is the set $\{ o_{\alpha_{\rm r}, R} \mid \alpha_{\rm r} \in (A_{\rm r})^{*}
\land R \subseteq A \}$, with the probability $p_{\alpha_{\rm r}, R}$ associated with $o_{\alpha_{\rm r},
R}$ being taken from the distribution whose values are of the form $1 / 2^{i}$, $i \in \natns_{> 0}$. Note
that $\calt$ is not finite state, but this affects only the initial step, whose only purpose is to
internally select a specific ready trace. \\
After this step, $\calt$ interacts with the process under test. Let $\rho \in (A \times 2^{A})^{*}$ be a
ready trace of the form $(a_{1}, R_{1}) \dots (a_{n}, R_{n})$, where $n \in \natns$. Given $s \in S$,
consider the trace $\alpha_{\rho, \rm r} \in (A_{\rm r})^{*}$ of length $n + 1$ in which the first element
is $a_{1} \triangleleft R$, with $R \subseteq A$ being the set of actions labeling the outgoing transitions
of $s$, the subsequent elements are of the form $a_{i} \triangleleft R_{i - 1}$ for $i = 2, \dots, n$, and
the last element is (i) $\circ \triangleleft \emptyset$ if $R_{n} = \emptyset$ or (ii) $\_ \triangleleft
R_{n}$ if $R_{n} \neq \emptyset$. Then for all $\calz \in \ms{Res}(s)$ it holds that:
\cws{0}{\hspace*{-0.8cm} \ms{prob}(\calrtcc(z_{s}, \rho)) \: = \: 0}
if there is no $a_{1} \dots a_{n}$-compatible computation from $z_{s}$, otherwise:
\cws{0}{\hspace*{-0.8cm} \ms{prob}(\calrtcc(z_{s}, \rho)) \: = \: \ms{prob}(\calscc(z_{s_{\rm r}, o},
\alpha_{\rho, \rm r})) / p_{\alpha'_{\rho, \rm r}, R_{n}}}
where $\alpha'_{\rho, \rm r}$ is $\alpha_{\rho, \rm r}$ without its last element. \\
Suppose that $s_{1} \sbis{\textrm{PTe-tbt,dis}} s_{2}$, which implies that $s_{1}$ and $s_{2}$ have the same
set $R$ of actions labeling their outgoing transitions and $s_{1, \rm r} \sbis{\textrm{PTe-tbt,dis}} s_{2,
\rm r}$. Then:

				\begin{itemize}

\item For each $\calz_{1} \in \ms{Res}(s_{1})$ there exists $\calz_{2} \in \ms{Res}(s_{2})$ such that for
all ready traces \linebreak $\rho = (a_{1}, R_{1}) \dots (a_{n}, R_{n}) \in (A \times 2^{A})^{*}$ either:
\cws{0}{\hspace*{-1.6cm} \ms{prob}(\calrtcc(z_{s_{1}}, \rho)) \: = \: 0 \: = \:
\ms{prob}(\calrtcc(z_{s_{2}}, \rho))}
or:
\cws{4}{\hspace*{-1.6cm}\begin{array}{rcccl}
\ms{prob}(\calrtcc(z_{s_{1}}, \rho)) & \!\!\! = \!\!\! & \ms{prob}(\calscc(z_{s_{1, \rm r}, o},
\alpha_{\rho, \rm r})) / p_{\alpha'_{\rho, \rm r}, R_{n}} & \!\!\! = \!\!\! & \\
& \!\!\! = \!\!\! & \ms{prob}(\calscc(z_{s_{2, \rm r}, o}, \alpha_{\rho, \rm r})) / p_{\alpha'_{\rho, \rm
r}, R_{n}} & \!\!\! = \!\!\! & \ms{prob}(\calrtcc(z_{s_{2}}, \rho)) \\
\end{array}}

\item Symmetrically for each $\calz_{2} \in \ms{Res}(s_{2})$.

				\end{itemize}

\noindent
This means that $s_{1} \sbis{\rm PRTr,dis} s_{2}$.

\item We preliminarily observe that for all $s \in S$, $\calz \in \ms{Res}(s)$, $n \in \natns$, $\alpha =
a_{1} \dots a_{n} \in A^{*}$, and $F_{1}, \dots, F_{n}, \linebreak R_{1}, \dots, R_{n} \in 2^{A}$ it holds
that:
\cws{0}{\hspace*{-0.8cm}\begin{array}{l}
\ms{prob}(\calftcc(z_{s}, (a_{1}, F_{1}) \dots (a_{n}, F_{n}))) \: = \\
\hspace*{3.0cm} = \: \sum_{R'_{1}, \dots, R'_{n} \in 2^{A} \, {\rm s.t.} \, R'_{i} \cap F_{i} = \emptyset \,
{\rm for \hspace{0.08cm} all} \, i = 1, \dots, n} \ms{prob}(\calrtcc(z_{s}, (a_{1}, R'_{1}) \dots (a_{n},
R'_{n}))) \\[0.1cm]
\ms{prob}(\calrtcc(z_{s}, (a_{1}, R_{1}) \dots (a_{n}, R_{n}))) \: = \: \ms{prob}(\calftcc(z_{s}, (a_{1},
\overline{R}_{1}) \dots (a_{n}, \overline{R}_{n}))) \\
\hspace*{3.0cm} - \sum_{R'_{1}, \dots, R'_{n} \in 2^{A} \, {\rm s.t.} \, R'_{i} \subseteq R_{i} \, {\rm for
\hspace{0.08cm} all} \, i = 1, \dots, n}^{R'_{j} \subset R_{J} \, {\rm for \hspace{0.08cm} some} \, j = 1,
\dots, n} \ms{prob}(\calrtcc(z_{s}, (a_{1}, R'_{1}) \dots (a_{n}, R'_{n}))) \\
\end{array}}
where $\overline{R}_{i} = A \setminus R_{i}$ for all $i = 1, \dots, n$. \\
Suppose that $s_{1} \sbis{\rm PRTr,dis} s_{2}$. Then we immediately derive that:

				\begin{itemize}

\item For each $\calz_{1} \in \ms{Res}(s_{1})$ there exists $\calz_{2} \in \ms{Res}(s_{2})$ such that for
all $(a_{1}, F_{1}) \dots (a_{n}, F_{n}) \in (A \times 2^{A})^{*}$:
\cws{0}{\hspace*{-1.6cm}\begin{array}{l}
\ms{prob}(\calftcc(z_{s_{1}}, (a_{1}, F_{1}) \dots (a_{n}, F_{n}))) \: = \\
\hspace*{1.7cm} = \: \sum_{R'_{1}, \dots, R'_{n} \in 2^{A} \, {\rm s.t.} \, R'_{i} \cap F_{i} = \emptyset \,
{\rm for \hspace{0.08cm} all} \, i = 1, \dots, n} \ms{prob}(\calrtcc(z_{s_{1}}, (a_{1}, R'_{1}) \dots
(a_{n}, R'_{n}))) \\[0.1cm]
\hspace*{1.7cm} = \: \sum_{R'_{1}, \dots, R'_{n} \in 2^{A} \, {\rm s.t.} \, R'_{i} \cap F_{i} = \emptyset \,
{\rm for \hspace{0.08cm} all} \, i = 1, \dots, n} \ms{prob}(\calrtcc(z_{s_{2}}, (a_{1}, R'_{1}) \dots
(a_{n}, R'_{n}))) \\[0.1cm]
\hspace*{1.7cm} = \: \ms{prob}(\calftcc(z_{s_{2}}, (a_{1}, F_{1}) \dots (a_{n}, F_{n}))) \\
\end{array}}

\item Symmetrically for each $\calz_{2} \in \ms{Res}(s_{2})$.

				\end{itemize}

This means that $s_{1} \sbis{\rm PFTr,dis} s_{2}$. \\
Suppose now that $s_{1} \sbis{\rm PFTr,dis} s_{2}$. For each ready trace $\rho \in (A \times 2^{A})^{*}$
including at least one infinite ready set, it trivially holds that for all $\calz_{1} \in \ms{Res}(s_{1})$
and $\calz_{2} \in \ms{Res}(s_{2})$:
\cws{0}{\hspace*{-0.8cm} \ms{prob}(\calrtcc(z_{s_{1}}, \rho)) \: = \: 0 \: = \:
\ms{prob}(\calrtcc(z_{s_{2}}, \rho))}
whenever the considered NPLTS is finitely branching. Thus, in order to prove that $s_{1} \sbis{\rm PRTr,dis}
s_{2}$, we can restrict ourselves to ready traces including only finite ready sets. Given an arbitrary
$\calz_{1} \in \ms{Res}(s_{1})$ that is matched by some $\calz_{2} \in \ms{Res}(s_{2})$ according to
$\sbis{\rm PFTr, dis}$, we show that the matching holds also under $\sbis{\rm PRTr,dis}$ by proceeding by
induction on the sum $k \in \natns$ of the cardinalities of the ready sets occurring in ready traces
including only finite ready sets:

				\begin{itemize}

\item Let $k = 0$, i.e., consider ready traces whose ready sets are all empty. Then for all $\alpha = a_{1}
\dots a_{n} \in A^{*}$:
\cws{4}{\hspace*{-1.6cm}\begin{array}{rcl}
\ms{prob}(\calrtcc(z_{s_{1}}, (a_{1}, \emptyset) \dots (a_{n}, \emptyset))) & \!\!\! = \!\!\! &
\ms{prob}(\calftcc(z_{s_{1}}, (a_{1}, A) \dots (a_{n}, A))) \\
& \!\!\! = \!\!\! & \ms{prob}(\calftcc(z_{s_{2}}, (a_{1}, A) \dots (a_{n}, A))) \\
& \!\!\! = \!\!\! & \ms{prob}(\calrtcc(z_{s_{2}}, (a_{1}, \emptyset) \dots (a_{n}, \emptyset))) \\
\end{array}}

\item Let $k \in \natns_{> 0}$ and suppose that the result holds for all ready traces for which the sum of
the cardinalities of the ready sets is less than $k$. Then for all $(a_{1}, R_{1}) \dots (a_{n}, R_{n}) \in
(A \times 2^{A})^{*}$ such that $\sum_{1 \le i \le n} |R_{i}| = k$:
\cws{4}{\hspace*{-1.6cm}\begin{array}{rcl}
\ms{prob}(\calrtcc(z_{s_{1}}, (a_{1}, R_{1}) \dots (a_{n}, R_{n}))) & \!\!\! = \!\!\! &
\ms{prob}(\calftcc(z_{s_{1}}, (a_{1}, \overline{R}_{1}) \dots (a_{n}, \overline{R}_{n}))) \\
& & \hspace*{-4.5cm} - \sum_{R'_{1}, \dots, R'_{n} \in 2^{A} \, {\rm s.t.} \, R'_{i} \subseteq R_{i} \, {\rm
for \hspace{0.08cm} all} \, i = 1, \dots, n}^{R'_{j} \subset R_{J} \, {\rm for \hspace{0.08cm} some} \, j =
1, \dots, n} \ms{prob}(\calrtcc(z_{s_{1}}, (a_{1}, R'_{1}) \dots (a_{n}, R'_{n}))) \\
& \!\!\! = \!\!\! & \ms{prob}(\calftcc(z_{s_{2}}, (a_{1}, \overline{R}_{1}) \dots (a_{n},
\overline{R}_{n}))) \\
& & \hspace*{-4.5cm} - \sum_{R'_{1}, \dots, R'_{n} \in 2^{A} \, {\rm s.t.} \, R'_{i} \subseteq R_{i} \, {\rm
for \hspace{0.08cm} all} \, i = 1, \dots, n}^{R'_{j} \subset R_{J} \, {\rm for \hspace{0.08cm} some} \, j =
1, \dots, n} \ms{prob}(\calrtcc(z_{s_{2}}, (a_{1}, R'_{1}) \dots (a_{n}, R'_{n}))) \\
& \!\!\! = \!\!\! & \ms{prob}(\calrtcc(z_{s_{2}}, (a_{1}, R_{1}) \dots (a_{n}, R_{n}))) \\
\end{array}}

				\end{itemize}

\noindent
A similar result holds also starting from an arbitrary $\calz_{2} \in \ms{Res}(s_{2})$ that is matched by
some \linebreak $\calz_{1} \in \ms{Res}(s_{1})$ according to $\sbis{\rm PFTr,dis}$. Therefore, we can
conclude that $s_{1} \sbis{\rm PRTr,dis} s_{2}$.

\item We preliminarily observe that for all $s \in S$, $\calz \in \ms{Res}(s)$, $\alpha \in A^{*}$, and $F,
R \in 2^{A}$ it holds that:
\cws{0}{\hspace*{-0.8cm}\begin{array}{rcl}
\ms{prob}(\calfcc(z_{s}, (\alpha, F))) & \!\!\! = \!\!\! & \sum_{R' \in 2^{A} \, {\rm s.t.} \, R' \cap F =
\emptyset} \ms{prob}(\calrcc(z_{s}, (\alpha, R'))) \\[0.1cm]
\ms{prob}(\calrcc(z_{s}, (\alpha, R))) & \!\!\! = \!\!\! & \ms{prob}(\calfcc(z_{s}, (\alpha, A \setminus
R))) - \sum_{R' \subset R} \ms{prob}(\calrcc(z_{s}, (\alpha, R'))) \\
\end{array}}
Suppose that $s_{1} \sbis{\rm PR,dis} s_{2}$. Then we immediately derive that:

				\begin{itemize}

\item For each $\calz_{1} \in \ms{Res}(s_{1})$ there exists $\calz_{2} \in \ms{Res}(s_{2})$ such that for
all $(\alpha, F) \in A^{*} \times 2^{A}$:
\cws{2}{\hspace*{-1.6cm}\begin{array}{rcl}
\ms{prob}(\calfcc(z_{s_{1}}, (\alpha, F))) & \!\!\! = \!\!\! & \sum_{R' \in 2^{A} \, {\rm s.t.} \, R' \cap F
= \emptyset} \ms{prob}(\calrcc(z_{s_{1}}, (\alpha, R'))) \\[0.1cm]
& \!\!\! = \!\!\! & \sum_{R' \in 2^{A} \, {\rm s.t.} \, R' \cap F = \emptyset} \ms{prob}(\calrcc(z_{s_{2}},
(\alpha, R'))) \\[0.1cm]
& \!\!\! = \!\!\! & \ms{prob}(\calfcc(z_{s_{2}}, (\alpha, F))) \\
\end{array}}

\item Symmetrically for each $\calz_{2} \in \ms{Res}(s_{2})$.

				\end{itemize}

This means that $s_{1} \sbis{\rm PF,dis} s_{2}$. \\
Suppose now that $s_{1} \sbis{\rm PF,dis} s_{2}$. For each ready pair $(\alpha, R) \in A^{*} \times 2^{A}$
such that $R$ is infinite, it trivially holds that for all $\calz_{1} \in \ms{Res}(s_{1})$ and $\calz_{2}
\in \ms{Res}(s_{2})$:
\cws{0}{\hspace*{-0.8cm} \ms{prob}(\calrcc(z_{s_{1}}, (\alpha, R))) \: = \: 0 \: = \:
\ms{prob}(\calrcc(z_{s_{2}}, (\alpha, R)))}
whenever the considered NPLTS is finitely branching. Thus, in order to prove that $s_{1} \sbis{\rm PR,dis}
s_{2}$, we can restrict ourselves to ready pairs whose ready set is finite. Given an arbitrary $\calz_{1}
\in \ms{Res}(s_{1})$ that is matched by some $\calz_{2} \in \ms{Res}(s_{2})$ according to $\sbis{\rm
PF,dis}$, we show that the matching holds also under $\sbis{\rm PR,dis}$ by proceeding by induction on the
cardinality $k \in \natns$ of the ready set of ready pairs whose ready set is finite:

				\begin{itemize}

\item Let $k = 0$, i.e., consider ready pairs whose ready set is empty. Then for all $\alpha \in A^{*}$:
\cws{4}{\hspace*{-1.6cm}\begin{array}{rcccl}
\ms{prob}(\calrcc(z_{s_{1}}, (\alpha, \emptyset))) & \!\!\! = \!\!\! & \ms{prob}(\calfcc(z_{s_{1}}, (\alpha,
A))) & \!\!\! = \!\!\! & \\
& \!\!\! = \!\!\! & \ms{prob}(\calfcc(z_{s_{2}}, (\alpha, A))) & \!\!\! = \!\!\! &
\ms{prob}(\calrcc(z_{s_{2}}, (\alpha, \emptyset))) \\
\end{array}}

\item Let $k \in \natns_{> 0}$ and suppose that the result holds for all ready pairs whose ready set has
cardinality less than $k$. Then for all $(\alpha, R) \in A^{*} \times 2^{A}$ such that $|R| = k$:
\cws{4}{\hspace*{-1.6cm}\begin{array}{rcl}
\ms{prob}(\calrcc(z_{s_{1}}, (\alpha, R))) & \!\!\! = \!\!\! & \ms{prob}(\calfcc(z_{s_{1}}, (\alpha, A
\setminus R))) - \sum_{R' \subset R} \ms{prob}(\calrcc(z_{s_{1}}, (\alpha, R'))) \\[0.1cm]
& \!\!\! = \!\!\! & \ms{prob}(\calfcc(z_{s_{2}}, (\alpha, A \setminus R))) - \sum_{R' \subset R}
\ms{prob}(\calrcc(z_{s_{2}}, (\alpha, R'))) \\[0.1cm]
& \!\!\! = \!\!\! & \ms{prob}(\calrcc(z_{s_{2}}, (\alpha, R))) \\
\end{array}}

				\end{itemize}

\noindent
A similar result holds also starting from an arbitrary $\calz_{2} \in \ms{Res}(s_{2})$ that is matched by
some \linebreak $\calz_{1} \in \ms{Res}(s_{1})$ according to $\sbis{\rm PF,dis}$. Therefore, we can conclude
that $s_{1} \sbis{\rm PR,dis} s_{2}$.

\item Suppose that $s_{1} \sbis{\rm PFTr,dis} s_{2}$. Then we immediately derive that:

				\begin{itemize}

\item For each $\calz_{1} \in \ms{Res}(s_{1})$ there exists $\calz_{2} \in \ms{Res}(s_{2})$ such that for
all $(a_{1} \dots a_{n}, F) \in A^{*} \times 2^{A}$:
\cws{0}{\hspace*{-1.6cm}\begin{array}{rcl}
\ms{prob}(\calfcc(z_{s_{1}}, (a_{1} \dots a_{n}, F))) & \!\!\! = \!\!\! & \ms{prob}(\calftcc(z_{s_{1}},
(a_{1}, \emptyset) \dots (a_{n - 1}, \emptyset) (a_{n}, F))) \\
& \!\!\! = \!\!\! & \ms{prob}(\calftcc(z_{s_{2}}, (a_{1}, \emptyset) \dots (a_{n - 1}, \emptyset) (a_{n},
F))) \\
& \!\!\! = \!\!\! & \ms{prob}(\calfcc(z_{s_{2}}, (a_{1} \dots a_{n}, F))) \\
\end{array}}

\item Symmetrically for each $\calz_{2} \in \ms{Res}(s_{2})$.

				\end{itemize}

This means that $s_{1} \sbis{\rm PF,dis} s_{2}$.

\item Suppose that $s_{1} \sbis{\rm PFTr} s_{2}$. Then we immediately derive that for all $(a_{1} \dots
a_{n}, F) \in A^{*} \times 2^{A}$:

				\begin{itemize}

\item For each $\calz_{1} \in \ms{Res}(s_{1})$ there exists $\calz_{2} \in \ms{Res}(s_{2})$ such that:
\cws{0}{\hspace*{-1.6cm}\begin{array}{rcl}
\ms{prob}(\calfcc(z_{s_{1}}, (a_{1} \dots a_{n}, F))) & \!\!\! = \!\!\! & \ms{prob}(\calftcc(z_{s_{1}},
(a_{1}, \emptyset) \dots (a_{n - 1}, \emptyset) (a_{n}, F))) \\
& \!\!\! = \!\!\! & \ms{prob}(\calftcc(z_{s_{2}}, (a_{1}, \emptyset) \dots (a_{n - 1}, \emptyset) (a_{n},
F))) \\
& \!\!\! = \!\!\! & \ms{prob}(\calfcc(z_{s_{2}}, (a_{1} \dots a_{n}, F))) \\
\end{array}}

\item Symmetrically for each $\calz_{2} \in \ms{Res}(s_{2})$.

				\end{itemize}

This means that $s_{1} \sbis{\rm PF} s_{2}$.

\item Suppose that $s_{1} \sbis{\rm PFTr,\sqcup\sqcap} s_{2}$. Then we immediately derive that for all
$\varphi = (\alpha, F) \in A^{*} \times 2^{A}$:
\cws{0}{\hspace*{-0.8cm}\begin{array}{rcl}
\bigsqcup\limits_{\calz_{1} \in \ms{Res}_{\alpha}(s_{1})} \ms{prob}(\calfcc(z_{s_{1}}, \varphi)) & \!\!\! =
\!\!\! & \bigsqcup\limits_{\calz_{1} \in \ms{Res}_{\alpha}(s_{1})} \ms{prob}(\calftcc(z_{s_{1}}, (a_{1},
\emptyset) \dots (a_{n - 1}, \emptyset) (a_{n}, F))) \\[0.4cm]
& \!\!\! = \!\!\! & \bigsqcup\limits_{\calz_{2} \in \ms{Res}_{\alpha}(s_{2})} \ms{prob}(\calftcc(z_{s_{2}},
(a_{1}, \emptyset) \dots (a_{n - 1}, \emptyset) (a_{n}, F))) \\
& \!\!\! = \!\!\! & \bigsqcup\limits_{\calz_{2} \in \ms{Res}_{\alpha}(s_{2})} \ms{prob}(\calfcc(z_{s_{2}},
\varphi)) \\[0.4cm]
\bigsqcap\limits_{\calz_{1} \in \ms{Res}_{\alpha}(s_{1})} \ms{prob}(\calfcc(z_{s_{1}}, \varphi)) & \!\!\! =
\!\!\! & \bigsqcap\limits_{\calz_{1} \in \ms{Res}_{\alpha}(s_{1})} \ms{prob}(\calftcc(z_{s_{1}}, (a_{1},
\emptyset) \dots (a_{n - 1}, \emptyset) (a_{n}, F))) \\[0.4cm]
& \!\!\! = \!\!\! & \bigsqcap\limits_{\calz_{2} \in \ms{Res}_{\alpha}(s_{2})} \ms{prob}(\calftcc(z_{s_{2}},
(a_{1}, \emptyset) \dots (a_{n - 1}, \emptyset) (a_{n}, F))) \\
& \!\!\! = \!\!\! & \bigsqcap\limits_{\calz_{2} \in \ms{Res}_{\alpha}(s_{2})} \ms{prob}(\calfcc(z_{s_{2}},
\varphi)) \\
\end{array}}
where $a_{1} \dots a_{n} = \alpha$. This means that $s_{1} \sbis{\rm PF,\sqcup\sqcap} s_{2}$.
\fullbox

			\end{enumerate}

		\end{proof}

	\end{theorem}

	\begin{figure}[tp]

\input{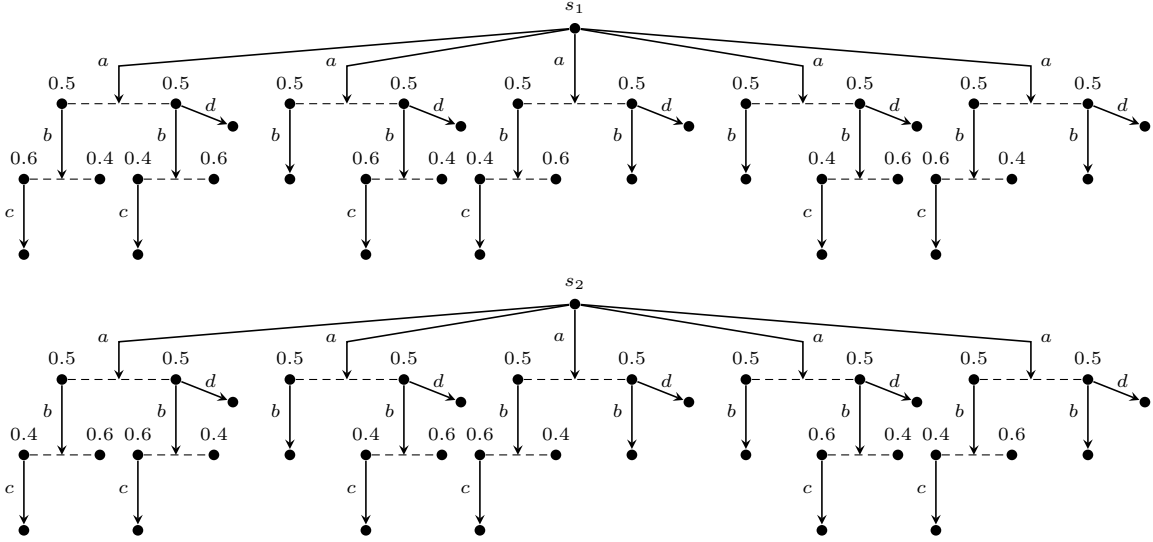}
\caption{Two NPLTS models distinguished by $\sbis{\rm PRTr,dis}$/$\sbis{\rm PFTr,dis}$ and identified by
$\sbis{\rm PR,dis}$/$\sbis{\rm PF,dis}$}
\label{fig:prtrdis_vs_prdis}

	\end{figure}

All the inclusions in Thm.~\ref{thm:decor_trace_results} are strict:

	\begin{itemize}

\item Figures~\ref{fig:dis_vs_by} and~\ref{fig:by_vs_supinf} respectively show that for all $\pi \in \{ {\rm
PRTr}, {\rm PFTr}, {\rm PR} \}$ it holds that $\sbis{\rm \pi,dis}$ is strictly finer than $\sbis{\rm \pi}$
and $\sbis{\rm \pi}$ is strictly finer than $\sbis{\rm \pi,\sqcup\sqcap}$.

\item Figure~\ref{fig:ptesupinf_anomaly} shows that $\sbis{\textrm{PTe-tbt,dis}}$ is strictly finer than
$\sbis{\rm PRTr,dis}$. It holds that $s_{1} \not\sbis{\textrm{PTe-tbt,dis}} s_{2}$ because
$\sbis{\textrm{PTe-tbt,dis}}$ coincides with $\sbis{\textrm{PTe-}\forall\exists}$ and the test in the
considered figure distinguishes the two processes with respect to $\sbis{\textrm{PTe-}\forall\exists}$. In
contrast, $s_{1} \sbis{\rm PRTr,dis} s_{2}$ because for each resolution of $s_{1}$ (resp.\ $s_{2}$) there
exists a resolution of $s_{2}$ (resp.\ $s_{1}$) having precisely the same ready-trace distribution.

\item Figure~\ref{fig:prtrdis_vs_prdis} shows that $\sbis{\rm PRTr,dis}$ and $\sbis{\rm PFTr,dis}$ are
strictly finer than $\sbis{\rm PR,dis}$ and $\sbis{\rm PF,dis}$, respectively. It holds that $s_{1}
\not\sbis{\rm PRTr,dis} s_{2}$ (and hence $s_{1} \not\sbis{\rm PFTr,dis} s_{2}$) because the ready-trace
distribution of the leftmost maximal resolution of $s_{1}$ in which the choice between $b$ and $d$ is
resolved in favor of $b$ is not matched by the ready-trace distribution of any of the resolutions of
$s_{2}$. In contrast, $s_{1} \sbis{\rm PR,dis} s_{2}$ (and hence $s_{1} \sbis{\rm PF,dis} s_{2}$) because
for each resolution of $s_{1}$ (resp.\ $s_{2}$) there exists a resolution of $s_{2}$ (resp.~$s_{1}$) having
precisely the same readiness distribution. In particular, the readiness distribution of the leftmost maximal
resolution of $s_{1}$ considered before is matched by the readiness distribution of the leftmost maximal
resolution of $s_{2}$ in which the choice between $b$ and $d$ is resolved as before, because when dealing
with ready pairs instead of ready traces the probabilities of performing the two $b$-transitions in those
resolutions can be summed up in the case of traces of length greater than~$1$. Likewise, the readiness
distribution of the leftmost maximal resolution of $s_{1}$ in which the choice between $b$ and $d$ is
resolved in favor of $d$ is matched by the readiness distribution of the central maximal resolution of
$s_{2}$ in which the choice between $b$ and $d$ is resolved in the same way.

\item Figure~\ref{fig:pftr_vs_pf} shows that $\sbis{\rm PFTr}$ and $\sbis{\rm PFTr,\sqcup\sqcap}$ are
strictly finer than $\sbis{\rm PF}$ and $\sbis{\rm PF,\sqcup\sqcap}$, respectively. It holds that $s_{1}
\not\sbis{\rm PFTr,\sqcup\sqcap} s_{2}$ (and hence $s_{1} \not\sbis{\rm PFTr} s_{2}$) because $s_{1}$ has a
computation compatible with the failure trace $(a, A \setminus \{ b, c \}) \, (c, A \setminus \{ e \}) \,
(e, A)$ while $s_{2}$ has no computation compatible with that failure trace. In contrast, $s_{1} \sbis{\rm
PF} s_{2}$ (and hence $s_{1} \sbis{\rm PF,\sqcup\sqcap} s_{2}$) because, given an arbitrary failure pair,
for each resolution of $s_{1}$ (resp.\ $s_{2}$) there exists a resolution of $s_{2}$ (resp.\ $s_{1}$) having
the same probability of performing a computation compatible with that failure pair.

	\end{itemize}

	\begin{figure}[tp]

\input{Pictures/pftr_vs_pf}
\caption{Two NPLTS models distinguished by $\sbis{\rm PFTr}$/$\sbis{\rm PFTr,\sqcup\sqcap}$ and identified
by $\sbis{\rm PF}$/$\sbis{\rm PF,\sqcup\sqcap}$}
\label{fig:pftr_vs_pf}

	\end{figure}

\pagebreak

Moreover:

	\begin{itemize}

\item $\sbis{\rm PB}$ and $\sbis{\rm PB,\sqcup\sqcap}$ are incomparable with the nine decorated-trace
equivalences introduced in this section. Indeed, in Fig.~\ref{fig:pb_pbsupinf_vs_others} it holds that
$s_{1} \sbis{\rm PB} s_{2}$ (and hence $s_{1} \sbis{\rm PB,\sqcup\sqcap} s_{2}$) -- as can be seen by taking
the equivalence relation that pairs states having equally labeled transitions leading to the same
distribution -- $s_{1} \not\sbis{\rm PRTr,\sqcup\sqcap} s_{2}$ (and hence $s_{1} \not\sbis{\rm PRTr} s_{2}$
and $s_{1} \not\sbis{\rm PRTr,dis} s_{2}$) -- due to the ready trace $(a, \{ b \}) \, (b, \{ c \}) \, (c,
\emptyset)$ having maximum probability $0.68$ in the first process and $0.61$ in the second process --
$s_{1} \not\sbis{\rm PR,\sqcup\sqcap} s_{2}$ (and hence $s_{1} \not\sbis{\rm PR} s_{2}$ and $s_{1}
\not\sbis{\rm PR,dis} s_{2}$) -- due to the ready pair $(a \, b \, c, \emptyset)$ having maximum probability
$0.68$ in the first process and $0.61$ in the second process -- and $s_{1} \not\sbis{\rm PFTr,\sqcup\sqcap}
s_{2}$ (and hence $s_{1} \not\sbis{\rm PFTr} s_{2}$ and $s_{1} \not\sbis{\rm PFTr,dis} s_{2}$) -- due to the
failure trace $(a, A \setminus \{ b \}) \, (b, A \setminus \{ c \}) \, (c, A)$ having maximum probability
$0.68$ in the first process and $0.61$ in the second process. In contrast, in
Fig.~\ref{fig:pbdis_vs_ptetbtdis} it holds that $s_{1} \not\sbis{\rm PB,\sqcup\sqcap} s_{2}$ (and hence
$s_{1} \not\sbis{\rm PB} s_{2}$) -- as the leftmost state with outgoing $b$-transitions reachable from
$s_{2}$ is not $\sqcup\sqcap$-bisimilar to the two states with outgoing $b$-transitions reachable from
$s_{1}$ -- and $s_{1} \sbis{\textrm{PTe-tbt,dis}} s_{2}$ (and hence $s_{1}$ and $s_{2}$ are also identified
by the nine decorated-trace equivalences).

\item $\sbis{\textrm{PTe-}\sqcup\sqcap}$ is incomparable with the nine decorated-trace equivalences
introduced in this section. Indeed, in Fig.~\ref{fig:by_vs_supinf} it holds that $s_{1}
\sbis{\textrm{PTe-}\sqcup\sqcap} s_{2}$, $s_{1} \not\sbis{\rm PRTr} s_{2}$ (and hence $s_{1} \not\sbis{\rm
PRTr,dis}$), $s_{1} \not\sbis{\rm PR} s_{2}$ (and hence $s_{1} \not\sbis{\rm PR,dis}$), and $s_{1}
\not\sbis{\rm PFTr} s_{2}$ (and hence $s_{1} \not\sbis{\rm PFTr,dis}$). In contrast, in
Fig.~\ref{fig:ptesupinf_anomaly} it holds that \linebreak $s_{1} \not\sbis{\textrm{PTe-}\sqcup\sqcap} s_{2}$
and $s_{1} \sbis{\rm PRTr,dis} s_{2}$ (and hence $s_{1} \sbis{\rm PR,dis} s_{2}$, $s_{1} \sbis{\rm PFTr,dis}
s_{2}$, $s_{1} \sbis{\rm PRTr} s_{2}$, $s_{1} \sbis{\rm PR} s_{2}$, and $s_{1} \sbis{\rm PFTr} s_{2}$).
Likewise, in Fig.~\ref{fig:pfsupinf_vs_ptesupinf} it holds that $s_{1} \sbis{\textrm{PTe-}\sqcup\sqcap}
s_{2}$ -- as there is no test that results in an interaction system having a maximal resolution with
differently labeled successful computations of the same length and hence no possibility of summing up their
success probabilities -- $s_{1} \not\sbis{\rm PRTr,\sqcup\sqcap} s_{2}$ -- due to the ready trace $(a, \{ b
\}) \, (b, \emptyset)$ whose maximum probability is $0.24$ in the first process and $0.21$ in the second
process -- $s_{1} \not\sbis{\rm PR,\sqcup\sqcap} s_{2}$ -- due to the ready pair $(a \, b, \emptyset)$ whose
maximum probability is $0.24$ in the first process and $0.21$ in the second process -- and $s_{1}
\not\sbis{\rm PFTr,\sqcup\sqcap} s_{2}$ -- due to the failure trace $(a, A \setminus \{ b \}) \, (b, A)$
whose maximum probability is $0.24$ in the first process and $0.21$ in the second process. In contrast, in
Fig.~\ref{fig:dis_vs_by} it holds that $s_{1} \not\sbis{\textrm{PTe-}\sqcup\sqcap} s_{2}$, $s_{1} \sbis{\rm
PRTr,\sqcup\sqcap} s_{2}$, $s_{1} \sbis{\rm PR,\sqcup\sqcap} s_{2}$, and $s_{1} \sbis{\rm PFTr,\sqcup\sqcap}
s_{2}$.

	\begin{figure}[tp]

\input{Pictures/pctrdis_vs_prtr}
\caption{Two NPLTS models distinguished by $\sbis{\rm PTr,\sqcup\sqcap}$ and identified by $\sbis{\rm
PRTr}$/$\sbis{\rm PR}$}
\label{fig:pctrdis_vs_prtr}

	\end{figure}

	\begin{figure}[tp]

\input{Pictures/prtr_vs_pctrdis}
\caption{Two NPLTS models distinguished by $\sbis{\rm PRTr,\sqcup\sqcap}$/$\sbis{\rm PR,\sqcup\sqcap}$ and
identified by $\sbis{\rm PCTr,dis}$/$\sbis{\rm PFTr}$}
\label{fig:prtr_vs_pctrdis}

	\end{figure}

\item $\sbis{\rm PRTr}$, $\sbis{\rm PR}$, $\sbis{\rm PRTr,\sqcup\sqcap}$, and $\sbis{\rm PR,\sqcup\sqcap}$
are incomparable with $\sbis{\rm PCTr,dis}$, $\sbis{\rm PTr,dis}$, $\sbis{\rm PFTr}$, $\sbis{\rm PF}$,
$\sbis{\textrm{PTe-tbt}}$, $\sbis{\rm PCTr}$, $\sbis{\rm PTr}$, $\sbis{\rm PFTr,\sqcup\sqcap}$, $\sbis{\rm
PF,\sqcup\sqcap}$, $\sbis{\textrm{PTe-tbt,$\sqcup\sqcap$}}$, $\sbis{\rm PCTr,\sqcup\sqcap}$, and $\sbis{\rm
PTr,\sqcup\sqcap}$. Indeed, in Fig.~\ref{fig:pctrdis_vs_prtr} it holds that $s_{1} \sbis{\rm PRTr} s_{2}$
(and hence $s_{1} \sbis{\rm PRTr,\sqcup\sqcap} s_{2}$), $s_{1} \sbis{\rm PR} s_{2}$ (and hence $s_{1}
\sbis{\rm PR,\sqcup\sqcap} s_{2}$), and $s_{1} \not\sbis{\rm PTr,\sqcup\sqcap} s_{2}$ (and hence $s_{1}
\not\sbis{\rm PCTr,\sqcup\sqcap} s_{2}$, $s_{1} \not\sbis{\textrm{PTe-tbt,$\sqcup\sqcap$}} s_{2}$, $s_{1}
\not\sbis{\rm PF,\sqcup\sqcap} s_{2}$, $s_{1} \not\sbis{\rm PFTr,\sqcup\sqcap} s_{2}$, $s_{1} \not\sbis{\rm
PTr} s_{2}$, $s_{1} \not\sbis{\rm PCTr} s_{2}$, $s_{1} \not\sbis{\textrm{PTe-tbt}} s_{2}$, $s_{1}
\not\sbis{\rm PF} s_{2}$, $s_{1} \not\sbis{\rm PFTr} s_{2}$, $s_{1} \not\sbis{\rm PTr,dis} s_{2}$, and
$s_{1} \not\sbis{\rm PCTr,dis} s_{2}$) -- due to the trace $a \, b$ having maximum probability $1$ in the
first process and $0.5$ in the second process. In contrast, in Fig.~\ref{fig:prtr_vs_pctrdis} it holds that
$s_{1} \not\sbis{\rm PRTr,\sqcup\sqcap} s_{2}$ (and hence $s_{1} \not\sbis{\rm PRTr} s_{2}$), $s_{1}
\not\sbis{\rm PR,\sqcup\sqcap} s_{2}$ (and hence $s_{1} \not\sbis{\rm PR} s_{2}$), $s_{1} \sbis{\rm
PCTr,dis} s_{2}$ (and hence $s_{1} \sbis{\rm PTr,dis} s_{2}$), and $s_{1} \sbis{\rm PFTr} s_{2}$ (and hence
$s_{1} \sbis{\rm PF} s_{2}$, $s_{1} \sbis{\textrm{PTe-tbt}} s_{2}$, $s_{1} \sbis{\rm PCTr} s_{2}$, $s_{1}
\sbis{\rm PTr} s_{2}$, $s_{1} \sbis{\rm PFTr,\sqcup\sqcap} s_{2}$, $s_{1} \sbis{\rm PF,\sqcup\sqcap} s_{2}$,
$s_{1} \sbis{\textrm{PTe-tbt,$\sqcup\sqcap$}} s_{2}$, $s_{1} \sbis{\rm PCTr,\sqcup\sqcap} s_{2}$, and $s_{1}
\sbis{\rm PTr,\sqcup\sqcap} s_{2}$).

\item $\sbis{\rm PRTr}$ and $\sbis{\rm PRTr,\sqcup\sqcap}$ are incomparable with $\sbis{\rm PR}$ and
$\sbis{\rm PR,\sqcup\sqcap}$. Indeed, in Fig.~\ref{fig:pr_vs_prtr} it holds that \linebreak $s_{1} \sbis{\rm
PRTr} s_{2}$ (and hence $s_{1} \sbis{\rm PRTr,\sqcup\sqcap} s_{2}$) and $s_{1} \not\sbis{\rm
PR,\sqcup\sqcap} s_{2}$ (and hence $s_{1} \not\sbis{\rm PR} s_{2}$) -- due to the ready pair $(a \, b \, f,
\emptyset)$ having maximum probability $1$ in the first process and $0.5$ in the second process. In
contrast, in Fig.~\ref{fig:pftr_vs_pf} it holds that $s_{1} \not\sbis{\rm PRTr,\sqcup\sqcap} s_{2}$ (and
hence $s_{1} \not\sbis{\rm PRTr} s_{2}$) -- due to the ready trace $(a, \{ b, c \}) \, (c, \{ e \}) \, (e,
\emptyset)$ having maximum probability $1$ in the first process and $0$ in the second process -- and $s_{1}
\sbis{\rm PR} s_{2}$ (and hence $s_{1} \sbis{\rm PR,\sqcup\sqcap} s_{2}$).

\item $\sbis{\rm PFTr}$ and $\sbis{\rm PFTr,\sqcup\sqcap}$ are incomparable with $\sbis{\rm PCTr,dis}$ and
$\sbis{\rm PTr,dis}$, because in Fig.~\ref{fig:dis_vs_by} it holds that $s_{1} \sbis{\rm PFTr} s_{2}$ (and
hence $s_{1} \sbis{\rm PFTr,\sqcup\sqcap} s_{2}$) and $s_{1} \not\sbis{\rm PTr,dis} s_{2}$ (and hence $s_{1}
\not\sbis{\rm PCTr,dis} s_{2}$), while in Fig.~\ref{fig:pf_vs_ptr} it holds that $s_{1} \not\sbis{\rm
PFTr,\sqcup\sqcap} s_{2}$ (and hence $s_{1} \not\sbis{\rm PFTr} s_{2}$) and $s_{1} \sbis{\rm PCTr,dis}
s_{2}$ (and hence $s_{1} \sbis{\rm PTr,dis} s_{2}$).

\item $\sbis{\rm PFTr,\sqcup\sqcap}$ is incomparable with $\sbis{\rm PCTr}$ and $\sbis{\rm PTr}$, because in
Fig.~\ref{fig:by_vs_supinf} it holds that $s_{1} \sbis{\rm PFTr,\sqcup\sqcap} s_{2}$ and $s_{1}
\not\sbis{\rm PTr} s_{2}$ (and hence $s_{1} \not\sbis{\rm PCTr} s_{2}$), while in Fig.~\ref{fig:pf_vs_ptr}
it holds that $s_{1} \not\sbis{\rm PFTr,\sqcup\sqcap} s_{2}$ and $s_{1} \sbis{\rm PCTr} s_{2}$ (and hence
$s_{1} \sbis{\rm PTr} s_{2}$).

\item $\sbis{\rm PFTr,\sqcup\sqcap}$ is incomparable with $\sbis{\rm PF}$ and $\sbis{\textrm{PTe-tbt}}$ too.
Indeed, in Fig.~\ref{fig:by_vs_supinf} it holds that $s_{1} \sbis{\rm PFTr,\sqcup\sqcap} s_{2}$ and $s_{1}
\not\sbis{\textrm{PTe-tbt}} s_{2}$ (and hence $s_{1} \not\sbis{\rm PF} s_{2}$). In contrast, in
Fig.~\ref{fig:pftr_vs_pf} it holds that $s_{1} \not\sbis{\rm PFTr,\sqcup\sqcap} s_{2}$ and $s_{1} \sbis{\rm
PF} s_{2}$. Likewise, in Fig.~\ref{fig:pfsupinf_vs_ptetbtsupinf} it holds that $s_{1} \not\sbis{\rm
PFTr,\sqcup\sqcap} s_{2}$ and $s_{1} \sbis{\textrm{PTe-tbt}} s_{2}$.

	\end{itemize}

	\begin{figure}[tp]

\input{Pictures/pr_vs_prtr}
\caption{Two NPLTS models distinguished by $\sbis{\rm PR}$/$\sbis{\rm PR,\sqcup\sqcap}$ and identified by
$\sbis{\rm PRTr}$/$\sbis{\rm PRTr,\sqcup\sqcap}$}
\label{fig:pr_vs_prtr}

	\end{figure}

%
\subsection{Simulation, Completed-Simulation, Failure-Simulation, and Ready-Simulation Equivalences}
\label{sec:sim_equiv}
%

\noindent
The variant of bisimulation equivalence in which only one direction is considered is called simulation
preorder, which is a refinement of trace inclusion. Simulation equivalence is defined as the kernel of
simulation preorder. In the probabilistic setting, simulation equivalence was defined by means of weight
functions in~\cite{JL91}. Here we shall follow an alternative characterization introduced in~\cite{DGJP03},
which relies on preorders as well as on closed sets. Given an NPLTS $(S, A, \! \arrow{}{} \!)$, a relation
$\cals$ over $S$, and $S' \subseteq S$, we say that $S'$ is an \emph{$\cals$-closed set} iff $\cals(S') = \{
s'' \in S \mid \exists s' \in S' \ldotp (s', s'') \in \cals \}$ is contained in $S'$. Notice that, if
$\cals$ is an equivalence relation, then an $\cals$-closed set is a group of equivalence classes.

	\begin{definition}\label{def:psdis}

(\emph{Probabilistic set-distribution similarity} -- $\sbis{\rm PS,dis}$ -- \cite{SL94}) \\
$s_{1} \sbis{\rm PS,dis} s_{2}$ iff $s_{1} \pre{\rm PS,dis} s_{2}$ and $s_{2} \pre{\rm PS,dis} s_{1}$, where
$\pre{\rm PS,dis}$ is the largest probabilistic set-distribution simulation. A preorder $\cals$ over $S$ is
a \emph{probabilistic set-distribution simulation} iff, whenever $(s_{1}, s_{2}) \in \cals$, then for each
$s_{1} \arrow{a}{} \cald_{1}$ there exists $s_{2} \arrow{a}{} \cald_{2}$ such that \underline{for all
$\cals$-closed $S' \subseteq S$} it holds that $\cald_{1}(S') \le \cald_{2}(S')$.
\fullbox

	\end{definition}

	\begin{definition}\label{def:ps}

(\emph{Probabilistic similarity} -- $\sbis{\rm PS}$ -- \cite{TDZ11}) \\
$s_{1} \sbis{\rm PS} s_{2}$ iff $s_{1} \pre{\rm PS} s_{2}$ and $s_{2} \pre{\rm PS} s_{1}$, where $\pre{\rm
PS}$ is the largest probabilistic simulation. A preorder $\cals$ over~$S$ is a \emph{probabilistic
simulation} iff, whenever $(s_{1}, s_{2}) \in \cals$, then \underline{for all $\cals$-closed $S' \subseteq
S$} it holds that for each $s_{1} \arrow{a}{} \cald_{1}$ there exists $s_{2} \arrow{a}{} \cald_{2}$ such
that $\cald_{1}(S') \le \cald_{2}(S')$.
\fullbox

	\end{definition}

	\begin{definition}\label{def:pssup}

(\emph{Probabilistic $\sqcup$-similarity} -- $\sbis{\rm PS,\sqcup}$) \\
$s_{1} \sbis{\rm PS,\sqcup} s_{2}$ iff $s_{1} \pre{\rm PS,\sqcup} s_{2}$ and $s_{2} \pre{\rm PS,\sqcup}
s_{1}$, where $\pre{\rm PS,\sqcup}$ is the largest probabilistic $\sqcup$-simulation. A preorder $\cals$
over $S$ is a \emph{probabilistic $\sqcup$-simulation} iff, whenever $(s_{1}, s_{2}) \in \cals$, then for
all $\cals$-closed $S' \subseteq S$ and $a \in A$ it holds that $s_{1} \arrow{a}{} \!$ implies $s_{2}
\arrow{a}{} \!$ and:
\cws{10}{\begin{array}{c}
\bigsqcup\limits_{s_{1} \arrow{a}{} \cald_{1}} \cald_{1}(S') \: \le \: \bigsqcup\limits_{s_{2} \arrow{a}{}
\cald_{2}} \cald_{2}(S') \\
\end{array}}
\fullbox

	\end{definition}

Similar to trace semantics, a number of variants of simulation semantics can be defined in which the sets of
actions that can be refused or accepted by states are also considered. Given $s \in S$, in the following we
let $\ms{init}(s) = \{ a \in A \mid s \arrow{a}{} \! \}$. Observing that $\ms{init}(s_{1}) \subseteq
\ms{init}(s_{2})$ whenever $s_{1}$ and $s_{2}$ are related by a simulation semantics, the additional
constraints are the following, where the names of the obtained variants are reported in parentheses:

	\begin{itemize}

\item $\ms{init}(s_{1}) = \emptyset \Longrightarrow \ms{init}(s_{2}) = \emptyset$, for \emph{completed
simulation} ($\sbis{\rm PCS,dis}$, $\sbis{\rm PCS}$, $\sbis{\rm PCS,\sqcup}$).

\item $\ms{init}(s_{1}) \cap F = \emptyset$ $\Longrightarrow$ $\ms{init}(s_{2}) \cap F = \emptyset$ for all
$F \in 2^{A}$, for \emph{failure simulation} ($\sbis{\rm PFS,dis}$, $\sbis{\rm PFS}$, $\sbis{\rm
PFS,\sqcup}$).

\item $\ms{init}(s_{1}) = \ms{init}(s_{2})$, for \emph{ready simulation} ($\sbis{\rm PRS,dis}$, $\sbis{\rm
PRS}$, $\sbis{\rm PRS,\sqcup}$).

	\end{itemize}

\noindent
Of the variants mentioned above, only $\sbis{\rm PFS,dis}$ has appeared in the literature of
nondeterministic and probabilistic processes~\cite{DGHM08,CR11}.

We now investigate the relationships of the twelve simulation-based equivalences among themselves and with
the various equivalences defined in~\cite{BDL13TCS} and in this paper. First of all, it turns out that every
simulation-based equivalence relying on partially matching transitions coincides with the corresponding
simulation-based equivalence relying on extremal probabilities. Moreover, ready-simulation semantics
coincides with failure-simulation semantics, but the various simulation-based semantics do not collapse to
bisimulation semantics as in the case of fully probabilistic processes~\cite{JL91}. Each of the
simulation-based equivalences relying on fully matching transitions is comprised between bisimilarity and
the corresponding trace equivalence, as in the fully nondeterministic spectrum~\cite{Gla01}. In contrast,
the simulation-based equivalences relying on partially matching transitions or extremal probabilities are
incomparable with most of the other equivalences.

	\begin{theorem}\label{thm:sim_results}

It holds that:

		\begin{enumerate}

\item $\sbis{\rm \pi,dis} \: \subseteq \: \sbis{\rm \pi} \: = \: \sbis{\rm \pi,\sqcup}$ for all $\pi \in \{
{\rm PS}, {\rm PCS}, {\rm PFS}, {\rm PRS} \}$ over image-finite NPLTS models.

\item $\sbis{\rm PB,\sigma'} \: \subseteq \: \sbis{\rm PRS,\sigma} \: = \: \sbis{\rm PFS,\sigma} \:
\subseteq \: \sbis{\rm PCS,\sigma} \: \subseteq \: \sbis{\rm PS,\sigma}$ for all $\sigma \in \{ {\rm dis},
\varepsilon, \sqcup \}$ and $\sigma' \in \{ {\rm dis}, \varepsilon, \sqcup\sqcap \}$ related to $\sigma$.

\item $\sbis{\rm PS,dis} \: \subseteq \: \sbis{\rm PTr,dis}$.

\item $\sbis{\rm PCS,dis} \: \subseteq \: \sbis{\rm PCTr,dis}$.

\item $\sbis{\rm PRS,dis} \: \subseteq \: \sbis{\textrm{PTe-tbt,dis}}$.

		\end{enumerate}

		\begin{proof}
Let $(S, A, \! \arrow{}{} \!)$ be an NPLTS and $s_{1}, s_{2} \in S$:

			\begin{enumerate}

\item The proof of the fact that $\sbis{\rm \pi,dis} \: \subseteq \: \sbis{\rm \pi} \: \subseteq \:
\sbis{\rm \pi,\sqcup}$ for all $\pi \in \{ {\rm PS}, {\rm PCS}, {\rm PFS}, {\rm PRS} \}$ is similar to the
proof of Thm.~6.5(1) in~\cite{BDL13TCS}. Moreover, it holds that $\sbis{\rm \pi,\sqcup} \: \subseteq \:
\sbis{\rm \pi}$ (and hence $\sbis{\rm \pi} \: = \: \sbis{\rm \pi,\sqcup}$) when the NPLTS is image finite.
In fact, supposing that $s_{1} \sbis{\rm \pi,\sqcup} s_{2}$, given a $\sbis{\rm \pi,\sqcup}$-closed set $S'
\subseteq S$ image finiteness guarantees that the following two sets:
\cws{0}{\hspace*{-0.8cm} \bigcup\limits_{s_{1} \arrow{a}{} \cald_{1}} \{ \cald_{1}(S') \} \quad \textrm{and}
\quad \bigcup\limits_{s_{2} \arrow{a}{} \cald_{2}} \{ \cald_{2}(S') \}}
are finite. In turn, the finiteness of those two sets ensures that their suprema respectively belong to the
two sets themselves. As a consequence, starting from:
\cws{0}{\hspace*{-0.8cm} \bigsqcup\limits_{s_{1} \arrow{a}{} \cald_{1}} \cald_{1}(S') \: \le \:
\bigsqcup\limits_{s_{2} \arrow{a}{} \cald_{2}} \cald_{2}(S')}
when both $s_{1}$ and $s_{2}$ have at least one outgoing $a$-transition, it holds that for each $s_{1}
\arrow{a}{} \cald'_{1}$ there exists $s_{2} \arrow{a}{} \cald'_{2}$ such that $\cald'_{1}(S') \le
\cald'_{2}(S')$ because we can take $\cald'_{2}$ such that $\cald'_{2}(S') = \bigsqcup_{s_{2} \arrow{a}{}
\cald_{2}} \cald_{2}(S')$. This means that $s_{1} \sbis{\rm \pi} s_{2}$.

\item The fact that $\sbis{\rm PB,\sigma'} \: \subseteq \: \sbis{\rm PRS,\sigma} \: \subseteq \: \sbis{\rm
PFS,\sigma} \: \subseteq \: \sbis{\rm PCS,\sigma} \: \subseteq \: \sbis{\rm PS,\sigma}$ for all $\sigma \in
\{ {\rm dis}, \varepsilon, \sqcup \}$ and $\sigma' \in \{ {\rm dis}, \varepsilon, \sqcup\sqcap \}$ related
to $\sigma$ is a straightforward consequence of the definition of the various equivalences. Moreover, it
holds that $\sbis{\rm PFS,\sigma} \: \subseteq \: \sbis{\rm PRS,\sigma}$ (and hence $\sbis{\rm PRS,\sigma}
\: = \: \sbis{\rm PFS,\sigma}$). In fact, supposing that $s_{1}$ and $s_{2}$ are related by a simulation
semantics so that $\ms{init}(s_{1}) \subseteq \ms{init}(s_{2})$, if $\ms{init}(s_{1}) \neq \ms{init}(s_{2})$
because of some $a \in A$ such that $a \notin \ms{init}(s_{1})$ and $a \in \ms{init}(s_{2})$ -- which means
that $s_{1} \not\sbis{\rm PRS,\sigma} s_{2}$ -- then $\ms{init}(s_{1}) \cap \{ a \} = \emptyset$ but
$\ms{init}(s_{2}) \cap \{ a \} \neq \emptyset$ -- which means that $s_{1} \not\sbis{\rm PFS,\sigma} s_{2}$.

\item We show that $s_{1} \pre{\rm PS,dis} s_{2} \Longrightarrow s_{1} \pre{\rm PTr,dis} s_{2}$ from which
the result will follow, where $s_{1} \pre{\rm PTr,dis} s_{2}$ means that for each $\calz_{1} \in
\ms{Res}(s_{1})$ there exists $\calz_{2} \in \ms{Res}(s_{2})$ such that for all $\alpha \in A^{*}$ it holds
that $\ms{prob}(\calcc(z_{s_{1}}, \alpha)) = \ms{prob}(\calcc(z_{s_{2}}, \alpha))$. \\
Suppose that $s_{1} \pre{\rm PS,dis} s_{2}$. This means that $(s_{1}, s_{2}) \in \cals$ for some
probabilistic set-distribution simulation $\cals$ over $S$. In turn, this induces projections of~$\cals$
that are fpr-simulations over pairs of matching resolutions and, since resolutions are fully probabilistic,
we derive from~\cite{JL91} that such projections are actually fpr-bisimulations~\cite{GJS90}.  As a
consequence, whenever $(r_{1}, r_{2}) \in \cals$, then for each $\calz_{1} \in \ms{Res}(r_{1})$ there exists
$\calz_{2} \in \ms{Res}(r_{2})$ such that the preorder $\cals_{1, 2}$ over $Z = Z_{1} \cup Z_{2}$
corresponding to $\cals$ projected onto $Z \times Z$ is an fpr-bisimulation, i.e., it is an equivalence
relation and, whenever $(z_{s'_{1}}, z_{s'_{2}}) \in \cals_{1, 2}$, then for each $z_{s'_{1}} \arrow{a}{}
\cald_{1}$ there exists $z_{s'_{2}} \arrow{a}{} \cald_{2}$ such that for all equivalence classes $C \in Z /
\cals_{1, 2}$ it holds that $\cald_{1}(C) = \cald_{2}(C)$. \\
Given $s'_{1}$, $s'_{2} \in S$ such that $(s'_{1}, s'_{2}) \in \cals$ and given $\calz_{1} \in
\ms{Res}(s'_{1})$ and $\calz_{2} \in \ms{Res}(s'_{2})$ such that $z_{s'_{1}}$ and $z_{s'_{2}}$ are related
by one of the projections of $\cals$, we prove that for all $\alpha \in A^{*}$ it holds that:
\cws{0}{\hspace*{-0.8cm} \ms{prob}(\calcc(z_{s'_{1}}, \alpha)) \: = \: \ms{prob}(\calcc(z_{s'_{2}},
\alpha))}
by proceeding by induction on the length $n$ of $\alpha$:

				\begin{itemize}

\item If $n = 0$, i.e., $\alpha = \varepsilon$, then:
\cws{4}{\hspace*{-1.6cm} \ms{prob}(\calcc(z_{s'_{1}}, \alpha)) \: = \: 1 \: = \:
\ms{prob}(\calcc(z_{s'_{2}}, \alpha))}

\item Let $n \in \natns_{> 0}$ and suppose that the result holds for all traces of length $m = 0, \dots, n -
1$ that label computations starting from pairs of states of $Z$ related by one of the projections
of~$\cals$. Assume that $\alpha = a \, \alpha'$. Given $s \in S$ and $\calz \in \ms{Res}(s)$, it holds that,
whenever $z_{s} \arrow{a}{} \cald$, then:
\cws{0}{\hspace*{-1.6cm} \ms{prob}(\calcc(z_{s}, \alpha)) \: = \: \sum\limits_{z_{s'} \in Z} \cald(z_{s'})
\cdot \ms{prob}(\calcc(z_{s'}, \alpha')) \: = \: \sum\limits_{[z_{s'}] \in Z / \cals'} \cald([z_{s'}]) \cdot
\ms{prob}(\calcc(z_{s'}, \alpha'))}
where $\cals'$ is a projection of $\cals$ and the factorization of $\ms{prob}(\calcc(z_{s'}, \alpha'))$ with
respect to the specific representative $z_{s'}$ of the equivalence class~$[z_{s'}]$ stems from the
application of the induction hypothesis on $\alpha'$ to all states of that equivalence class. Since
$z_{s'_{1}}$ and $z_{s'_{2}}$ are related by a projection $\cals_{1, 2}$ of $\cals$, it follows that,
whenever $z_{s'_{1}} \arrow{a}{} \cald_{1}$, then $z_{s'_{2}} \arrow{a}{} \cald_{2}$ and:
\cws{10}{\hspace*{-1.6cm}\begin{array}{rcccl}
\ms{prob}(\calcc(z_{s'_{1}}, \alpha)) & \!\!\! = \!\!\! & \sum\limits_{[z_{s'}] \in Z / \cals_{1, 2}}
\cald_{1}([z_{s'}]) \cdot \ms{prob}(\calcc(z_{s'}, \alpha')) & \!\!\! = \!\!\! & \\[0.4cm]
& \!\!\! = \!\!\! & \sum\limits_{[z_{s'}] \in Z / \cals_{1, 2}} \cald_{2}([z_{s'}]) \cdot
\ms{prob}(\calcc(z_{s'}, \alpha')) & \!\!\! = \!\!\! & \ms{prob}(\calcc(z_{s'_{2}}, \alpha)) \\
\end{array}}

				\end{itemize}

Therefore $s_{1} \pre{\rm PTr,dis} s_{2}$.

\item The proof that $s_{1} \pre{\rm PCS,dis} s_{2} \Longrightarrow s_{1} \pre{\rm PCTr,dis} s_{2}$, from
which the result follows, is similar to the proof of the previous result. We note that:

				\begin{itemize}

\item For fully probabilistic models like resolutions, fpr-completed simulations are coarser than \linebreak
fpr-bisimulations and finer than fpr-simulations. Since fpr-simulations are fpr-bisimulations over these
models~\cite{JL91}, also fpr-completed simulations are fpr-bisimulations.

\item In the base case of the induction, it additionally holds that:
\cws{0}{\hspace*{-1.6cm} \ms{prob}(\calccc(z_{s'_{1}}, \alpha)) \: = \: \ms{prob}(\calccc(z_{s'_{2}},
\alpha)) \: = \: \left\{ \begin{array}{ll}
1 & \hspace{0.5cm} \textrm{if $\ms{init}(s'_{1}) = \ms{init}(s'_{2}) = \emptyset$} \\
0 & \hspace{0.5cm} \textrm{if $\ms{init}(s'_{1}) \neq \emptyset \neq \ms{init}(s'_{2})$} \\
\end{array} \right.}
where $\ms{init}(s'_{1}) = \emptyset$ iff $\ms{init}(s'_{2}) = \emptyset$ because $(s'_{1}, s'_{2}) \in
\cals$ and $\cals$ is a probabilistic set-distribution completed simulation.

\item In the general case of the induction, $\ms{prob}(\calccc(z_{s}, \alpha))$ is expressed recursively in
the same way as $\ms{prob}(\calcc(z_{s}, \alpha))$.

				\end{itemize}

\item The proof that $s_{1} \pre{\rm PRS,dis} s_{2} \Longrightarrow s_{1} \pre{\textrm{PTe-tbt,dis}} s_{2}$,
from which the result follows, is similar to the proof of Thm.~6.5(2) in~\cite{BDL13TCS}. We note that:

				\begin{itemize}

\item We exploit the fact that states related by $\pre{\rm PRS,dis}$ have the same set of actions labeling
their outgoing transitions to establish a connection among resolutions of the interaction systems that are
maximal (remember that states not enjoying that property are trivially distinguished by
$\pre{\textrm{PTe-tbt,dis}}$).

\item For fully probabilistic models like maximal resolutions, fpr-ready simulations are coarser than
fpr-bisimulations and finer than fpr-simulations. Since fpr-simulations are fpr-bisimulations over these
models~\cite{JL91}, also fpr-ready simulations are fpr-bisimulations.
\fullbox

				\end{itemize}

			\end{enumerate}

		\end{proof}

	\end{theorem}

	\begin{figure}[tp]

\input{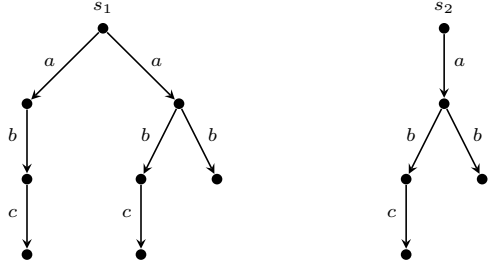}
\caption{Two NPLTS models distinguished by $\sbis{\rm PB,dis}$/$\sbis{\rm PB}$/$\sbis{\rm PB,\sqcup\sqcap}$
and identified by $\sbis{\rm PRS,dis}$/$\sbis{\rm PRS}$/$\sbis{\rm PRS,\sqcup}$}
\label{fig:pbdis_vs_prsdis}

	\end{figure}

All the inclusions in Thm.~\ref{thm:sim_results} are strict:

	\begin{itemize}

\item Figures~\ref{fig:dis_vs_by} and~\ref{fig:by_vs_supinf} respectively show that for all $\pi \in \{ {\rm
PS}, {\rm PCS}, {\rm PFS}, {\rm PRS} \}$ it holds that $\sbis{\rm \pi,dis}$ is strictly finer than
$\sbis{\rm \pi}$ and $\sbis{\rm \pi}$ is strictly finer than $\sbis{\rm \pi,\sqcup}$.

\item Figure~\ref{fig:pbdis_vs_prsdis} shows that $\sbis{\rm PB,\sigma'}$ is strictly finer than $\sbis{\rm
PRS,\sigma}$ for all $\sigma \in \{ {\rm dis}, \varepsilon, \sqcup \}$ and $\sigma' \in \{ {\rm dis},
\varepsilon, \sqcup\sqcap \}$ related to $\sigma$. In particular, $s_{1}$ and $s_{2}$ are not bisimilar
because the leftmost state with outgoing \linebreak $b$-transitions reachable from $s_{1}$ is not bisimilar
to the only state with outgoing $b$-transitions reachable from $s_{2}$.

\item Figure~\ref{fig:pf_vs_ptr} shows that $\sbis{\rm PRS,\sigma}$ is strictly finer than $\sbis{\rm
PCS,\sigma}$ for all $\sigma \in \{ {\rm dis}, \varepsilon, \sqcup \}$. In particular, $s_{1}$ and $s_{2}$
are not ready similar because the leftmost state with outgoing $b$-transitions reachable from $s_{1}$ is not
ready similar to the only state with outgoing $b$-transitions reachable from $s_{2}$.

\item Figure~\ref{fig:ptrsupinf_anomaly_maxres} shows that $\sbis{\rm PCS,\sigma}$ is strictly finer than
$\sbis{\rm PS,\sigma}$ for all $\sigma \in \{ {\rm dis}, \varepsilon, \sqcup \}$. In particular, $s_{1}$
and~$s_{2}$ are not completed similar because the rightmost state reachable from $s_{1}$ after performing
$a$ is not completed similar to the only state reachable from $s_{2}$ after performing $a$.

\item Figure~\ref{fig:pbdis_vs_ptetbtdis} shows that $\sbis{\rm PS,dis}$, $\sbis{\rm PCS,dis}$, and
$\sbis{\rm PRS,dis}$ are strictly finer than $\sbis{\rm PTr,dis}$, $\sbis{\rm PCTr,dis}$, and
$\sbis{\textrm{PTe-tbt,dis}}$, respectively. It holds that $s_{1} \not\sbis{\rm PS,dis} s_{2}$ (and hence
$s_{1} \not\sbis{\rm PCS,dis} s_{2}$ and $s_{1} \not\sbis{\rm PRS,dis} s_{2}$) because the leftmost state
with outgoing $b$-transitions reachable from $s_{2}$ is not set-distribution similar to the two states with
outgoing $b$-transitions reachable from $s_{1}$. In contrast, $s_{1} \sbis{\textrm{PTe-tbt,dis}} s_{2}$ (and
hence $s_{1} \sbis{\rm PCTr,dis} s_{2}$ and $s_{1} \sbis{\rm PTr,dis} s_{2}$) because success probabilities
are computed in a trace-by-trace fashion without adding up over different traces.

	\end{itemize}

\pagebreak

Moreover:

	\begin{itemize}

\item $\sbis{\rm PCS,dis}$ is incomparable with the five testing equivalences and the twelve decorated-trace
equivalences. Indeed, in Fig.~\ref{fig:pf_vs_ptr} it holds that $s_{1} \sbis{\rm PCS,dis} s_{2}$ -- as can
be seen by taking the preorder that pairs states having at least one equally labeled transition -- $s_{1}
\not\sbis{\textrm{PTe-tbt,$\sqcup\sqcap$}} s_{2}$ (and hence $s_{1}$ and $s_{2}$ are also distinguished by
the other four testing equivalences, the three failure equivalences, and the three failure-trace
equivalences) -- due to the test having an $a$-transition followed by a $c$-transition leading to success,
which results in a maximal resolution with completed trace $a$ when interacting with the first process and
no maximal resolution with completed trace $a$ when interacting with the second process -- and $s_{1}
\not\sbis{\rm PR,\sqcup\sqcap} s_{2}$ and $s_{1} \not\sbis{\rm PRTr,\sqcup\sqcap} s_{2}$ (and hence $s_{1}$
and $s_{2}$ are also distinguished by the other two readiness equivalences and the other two ready-trace
equivalences) -- due to the ready pair and ready trace $(a, \{ b \})$ having maximum probability $1$ in the
first process and $0$ in the second process. In contrast, in Fig.~\ref{fig:pbdis_vs_ptetbtdis} it holds that
$s_{1} \not\sbis{\rm PCS,dis} s_{2}$ -- as the leftmost state with outgoing $b$-transitions reachable from
$s_{2}$ is not set-distribution completed similar to the two states with outgoing $b$-transitions reachable
from $s_{1}$ -- and $s_{1} \sbis{\textrm{PTe-tbt,dis}} s_{2}$ (and hence $s_{1}$ and $s_{2}$ are also
identified by the other four testing equivalences and the twelve decorated-trace equivalences).

\item $\sbis{\rm PS,dis}$ is incomparable with the five testing equivalences, the twelve decorated-trace
equivalences, and the three completed-trace equivalences. Indeed, in Fig.~\ref{fig:ptrsupinf_anomaly_maxres}
it holds that $s_{1} \sbis{\rm PS,dis} s_{2}$ -- as can be seen by taking the preorder that pairs states
having equally labeled transitions -- $s_{1} \not\sbis{\rm PCTr,\sqcup\sqcap} s_{2}$ (and hence $s_{1}$ and
$s_{2}$ are also distinguished by the other two completed-trace equivalences, the three failure
equivalences, and the three failure-trace equivalences) -- due to the completed trace $a$ having maximum
probability $1$ in the first process and $0$ in the second process -- $s_{1}
\not\sbis{\textrm{PTe-tbt,$\sqcup\sqcap$}} s_{2}$ (and hence $s_{1}$ and $s_{2}$ are also distinguished by
the other four testing equivalences) -- due to the test having an $a$-transition followed by a
$b$-transition leading to success, which results in a maximal resolution with completed trace $a$ when
interacting with the first process and no maximal resolution with completed trace $a$ when interacting with
the second process -- and $s_{1} \not\sbis{\rm PR,\sqcup\sqcap} s_{2}$ and $s_{1} \not\sbis{\rm
PRTr,\sqcup\sqcap} s_{2}$ (and hence $s_{1}$ and $s_{2}$ are also distinguished by the other two readiness
equivalences and the other two ready-trace equivalences) -- due to the ready pair and ready trace $(a,
\emptyset)$ having maximum probability $1$ in the first process and $0$ in the second process. In contrast,
in Fig.~\ref{fig:pbdis_vs_ptetbtdis} it holds that $s_{1} \not\sbis{\rm PS,dis} s_{2}$ -- as the leftmost
state with outgoing $b$-transitions reachable from $s_{2}$ is not set-distribution similar to the two states
with outgoing $b$-transitions reachable from $s_{1}$ -- and $s_{1} \sbis{\textrm{PTe-tbt,dis}} s_{2}$ (and
hence $s_{1}$ and $s_{2}$ are also identified by the other four testing equivalences, the twelve
decorated-trace equivalences, and the three completed-trace equivalences).

\item $\sbis{\rm \pi}$ and $\sbis{\rm \pi,\sqcup}$ are incomparable with the five testing equivalences and
the eighteen trace-based equivalences for all $\pi \in \{ {\rm PS}, {\rm PCS}, {\rm PFS}, {\rm PRS} \}$.
Indeed, in Fig.~\ref{fig:pb_pbsupinf_vs_others} it holds that $s_{1} \sbis{\rm PRS} s_{2}$ (and hence
$s_{1}$ and $s_{2}$ are also identified by the other seven simulation-based equivalences) -- as can be seen
by taking the preorder that pairs states having equally labeled transitions leading to the same distribution
-- and $s_{1} \not\sbis{\rm PTr,\sqcup\sqcap} s_{2}$, $s_{1} \not\sbis{\rm PR,\sqcup\sqcap} s_{2}$, and
$s_{1} \not\sbis{\rm PRTr,\sqcup\sqcap} s_{2}$ (and hence $s_{1}$ and $s_{2}$ are also distinguished by the
other fifteen trace-based equivalences and the five testing equivalences) -- due to the trace $a \, b \, c$,
the ready pair $(a \, b \, c, \emptyset)$, and the ready trace $(a, \{ b \}) \, (b, \{ c \}) \, (c,
\emptyset)$ having maximum probability $0.68$ in the first process and $0.61$ in the second process. In
contrast, in Fig.~\ref{fig:pbdis_vs_ptetbtdis} it holds that $s_{1} \not\sbis{\rm PS,\sqcup\sqcap} s_{2}$
(and hence $s_{1}$ and~$s_{2}$ are also distinguished by the other seven simulation-based equivalences) --
as the leftmost state with outgoing $b$-transitions reachable from $s_{2}$ is not $\sqcup$-similar to the
two states with outgoing $b$-transitions reachable from $s_{1}$ -- and $s_{1} \sbis{\textrm{PTe-tbt,dis}}
s_{2}$ (and hence $s_{1}$ and $s_{2}$ are also identified by the other four testing equivalences and the
eighteen trace-based equivalences).

\item $\sbis{\rm PB}$ and $\sbis{\rm PB,\sqcup\sqcap}$ are incomparable with $\sbis{\rm \pi,dis}$ for all
$\pi \in \{ {\rm PS}, {\rm PCS}, {\rm PFS}, {\rm PRS} \}$, because in Fig.~\ref{fig:dis_vs_by} it holds that
$s_{1} \sbis{\rm PB} s_{2}$ (and hence $s_{1} \sbis{\rm PB,\sqcup\sqcap} s_{2}$) and $s_{1} \not\sbis{\rm
PS,dis} s_{2}$ (and hence $s_{1} \not\sbis{\rm PCS,dis} s_{2}$, $s_{1} \not\sbis{\rm PFS,dis} s_{2}$, and
$s_{1} \not\sbis{\rm PRS,dis} s_{2}$), while in Fig.~\ref{fig:pbdis_vs_prsdis} it holds that $s_{1}
\not\sbis{\rm PB,\sqcup\sqcap} s_{2}$ (and hence $s_{1} \not\sbis{\rm PB} s_{2}$) and $s_{1} \sbis{\rm
PRS,dis} s_{2}$ (and hence $s_{1} \sbis{\rm PFS,dis} s_{2}$, $s_{1} \sbis{\rm PCS,dis} s_{2}$, and $s_{1}
\sbis{\rm PS,dis} s_{2}$).

\item $\sbis{\rm PRS}$, $\sbis{\rm PFS}$, $\sbis{\rm PRS,\sqcup}$, and $\sbis{\rm PFS,\sqcup}$ are
incomparable with $\sbis{\rm PCS,dis}$ and $\sbis{\rm PS,dis}$, because in Fig.~\ref{fig:dis_vs_by} it holds
that $s_{1} \sbis{\rm PRS} s_{2}$ (and hence $s_{1} \sbis{\rm PFS} s_{2}$, $s_{1} \sbis{\rm PRS,\sqcup}
s_{2}$, and $s_{1} \sbis{\rm PFS,\sqcup} s_{2}$) and $s_{1} \not\sbis{\rm PS,dis} s_{2}$ (and hence $s_{1}
\not\sbis{\rm PCS,dis} s_{2}$), while in Fig.~\ref{fig:pf_vs_ptr} it holds that $s_{1} \not\sbis{\rm
PFS,\sqcup} s_{2}$ (and hence $s_{1} \not\sbis{\rm PRS,\sqcup} s_{2}$, $s_{1} \not\sbis{\rm PFS} s_{2}$, and
$s_{1} \not\sbis{\rm PRS} s_{2}$) and $s_{1} \sbis{\rm PCS,dis} s_{2}$ (and hence $s_{1} \sbis{\rm PS,dis}
s_{2}$).

\item $\sbis{\rm PCS}$ and $\sbis{\rm PCS,\sqcup}$ are incomparable with $\sbis{\rm PS,dis}$, because in
Fig.~\ref{fig:dis_vs_by} it holds that $s_{1} \sbis{\rm PCS} s_{2}$ (and hence $s_{1} \sbis{\rm PCS,\sqcup}
s_{2}$) and $s_{1} \not\sbis{\rm PS,dis} s_{2}$, while in Fig.~\ref{fig:ptrsupinf_anomaly_maxres} it holds
that $s_{1} \sbis{\rm PS,dis} s_{2}$ and $s_{1} \not\sbis{\rm PCS,\sqcup} s_{2}$ (and hence $s_{1}
\not\sbis{\rm PCS} s_{2}$).

	\end{itemize}

%
\subsection{A Full Spectrum}\label{sec:spectrum_full}
%

\begin{figure}[tp]

\centerline{\includegraphics{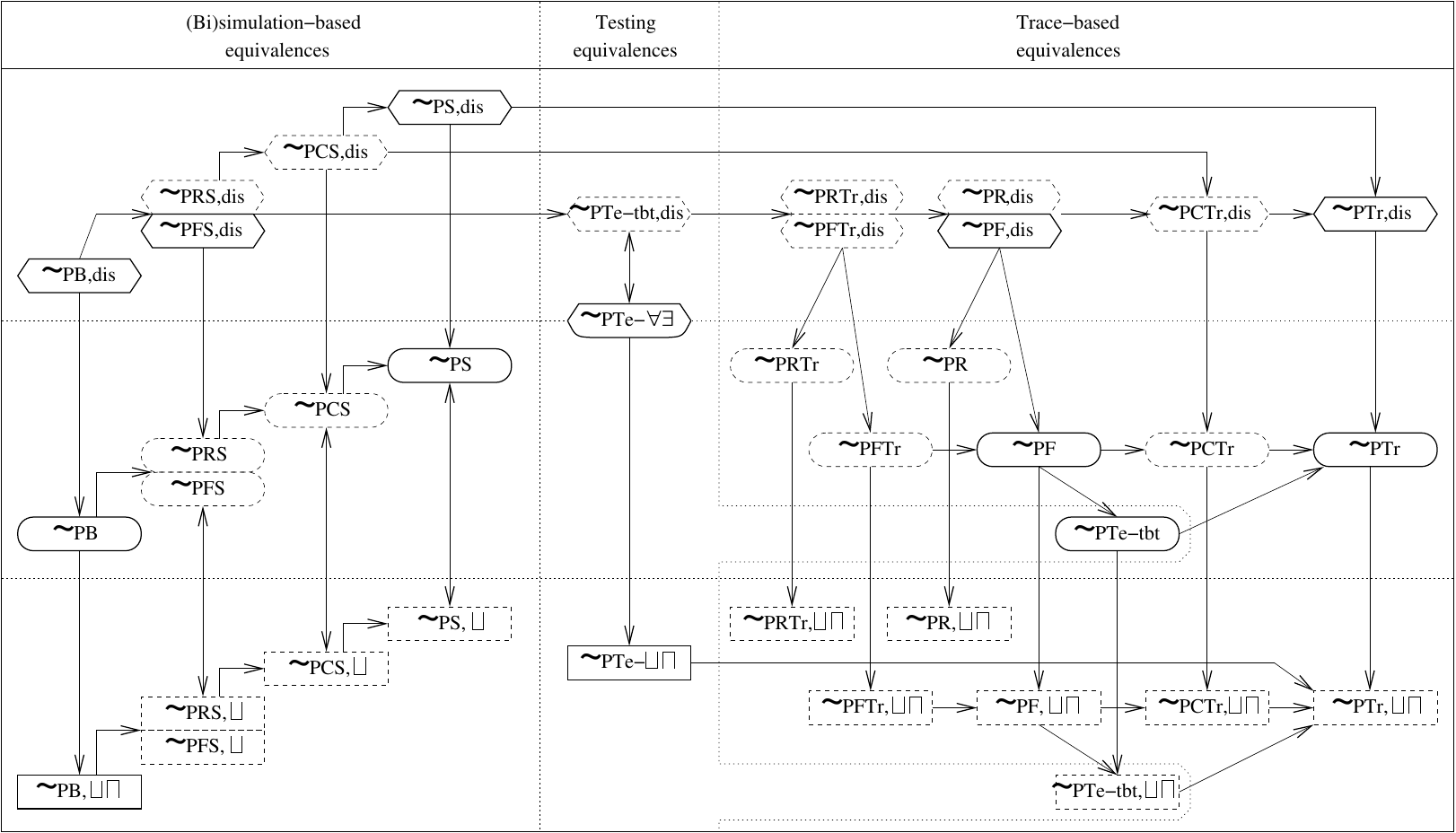}}
\caption{Full spectrum of strong behavioral equivalences for finitely-branching NPLTS models (deterministic
schedulers)}
\label{fig:spectrum_full}

	\end{figure}

\noindent
The spectum of all the considered equivalences is depicted in Fig.~\ref{fig:spectrum_full}. We have followed
the same graphical conventions mentioned at the beginning of Sect.~\ref{sec:spectrum}, with adjacency of
boxes within the same fragment having the same meaning as bidirectional arrows connecting boxes of different
fragments, i.e., coincidence. Note that there are many more dashed boxes (corresponding to equivalences
introduced in this paper) than in~Fig.~\ref{fig:spectrum_reduced}.

The top fragment of the spectrum in Fig.~\ref{fig:spectrum_full} refers to the considered equivalences that
are based on fully matching resolutions. Similar to the spectrum for fully probabilistic processes
in~\cite{JS90,HT92}, many equivalences collapse into a single one; in particular, ready-simulation semantics
coincides with failure-simulation semantics, ready-trace semantics coincides with failure-trace semantics,
and readiness semantics coincides with failure semantics. Different from the fully probabilistic spectrum,
in the top fragment we have that the various simulation-based semantics do not coincide with bisimulation
semantics~\cite{JL91} and that completed-trace semantics does not coincide with trace
semantics~\cite{JS90,HT92}. Moreover, testing semantics turns out to be finer than failure semantics.

The central fragment and the bottom fragment of the spectrum in Fig.~\ref{fig:spectrum_full} instead refer
to the considered equivalences that are based on partially matching resolutions and extremal probabilities,
respectively. These equivalences are coarser than those in the top fragment and do not flatten the
specificity of the intuition behind the original definition of the behavioral equivalences for LTS models.
Therefore, the two fragments at hand preserve much of the original spectrum in~\cite{Gla01} for fully
nondeterministic processes, with testing semantics being coarser than failure semantics. It is worth noting
the coincidence of corresponding simulation-based equivalences in the two fragments (due to the fact that
the comparison operator $\le$ is used in their definitions), whereas this is not the case for the two
bisimulation equivalences (as the comparison operator $=$ is used instead in their definitions). We finally
stress the isolation of bisimulation semantics, simulation semantics, ready-trace semantics, and readiness
semantics in the two fragments, as well as the partial isolation of $\sbis{\textrm{PTe-}\sqcup\sqcap}$.

%
%
\section{Deterministic Schedulers vs.\ Randomized Schedulers}\label{sec:det_vs_rand_sched}
%
%

\noindent
So far, we have considered strong equivalences for NPLTS models that compare probabilities calculated after
resolving nondeterminism by means of deterministic schedulers. We now examine the case of randomized
schedulers. Each of them selects at each state a convex combination of equally labeled transitions, which is
called a \emph{combined transition}~\cite{Seg95a}. Notice that a deterministic scheduler is a special case
of randomized scheduler in which every selected combination involves a single ordinary transition.

	\begin{definition}\label{def:resolution_rand}

Let $\call = (S, A, \! \arrow{}{} \!)$ be an NPLTS and $s \in S$. We say that an NPLTS $\calz = (Z, A, \!
\arrow{}{\calz} \!)$ is a \emph{resolution} of~$s$ obtained \emph{via a randomized scheduler} iff there
exists a state correspondence function $\ms{corr}_{\calz} : Z \rightarrow S$ such that $s =
\ms{corr}_{\calz}(z_{s})$, for some $z_{s} \in Z$, and for all $z \in Z$ it holds that:

		\begin{itemize}

\item If $z \arrow{a}{\calz} \cald$, then there exist $n \in \natns_{> 0}$, $\{ p_{i} \in \realns_{]0, 1]}
\mid 1 \le i \le n \}$, and $\{ \ms{corr}_{\calz}(z) \arrow{a}{} \cald_{i} \mid 1 \le i \le n \}$ such that
$\sum_{i = 1}^{n} p_{i} = 1$ and $\cald(z') = \sum_{i = 1}^{n} p_{i} \cdot \cald_{i}(\ms{corr}_{\calz}(z'))$
for all $z' \in Z$.

\item If $z \arrow{a_{1}}{\calz} \cald_{1}$ and $z \arrow{a_{2}}{\calz} \cald_{2}$, then $a_{1} = a_{2}$ and
$\cald_{1} = \cald_{2}$.
\fullbox

		\end{itemize}

	\end{definition}

For each strong behavioral equivalence $\sbis{}$ introduced in~\cite{BDL13TCS} and in this paper, we denote
by $\sbis{}^{\rm ct}$ the corresponding equivalence based on combined transitions (ct-equivalence for
short), i.e., in which nondeterminism is resolved by means of randomized schedulers. For the eighteen
trace-based equivalences and the five testing equivalences, the only modification in their definitions is
the use of $\ms{Res}^{\rm ct}$ in place of $\ms{Res}$, where $\ms{Res}^{\rm ct}$ is the set of resolutions
of a state obtained via a randomized scheduler. For the fifteen (bi)simulation-based equivalences, the only
modification in their definitions is the direct use of combined transitions (denoted by $\! \arrow{}{\rm c}
\!$) instead of ordinary transitions.

All the results connecting the various equivalences and the counterexamples showing strict inclusion or
incomparability are still valid for the ct-equivalences. A notable exception is given by the counterexamples
based on Fig.~\ref{fig:by_vs_supinf}, as the central $\ms{offer}$-transition of $s_{1}$ can now be obtained
as a convex combination of the two $\ms{offer}$-transitions of~$s_{2}$ with both coefficients equal to
$0.5$. Indeed, no ct-equivalence can be finer than the corresponding equivalence arising from deterministic
schedulers, as matching ordinary transitions induce matching combined transitions.

While every ct-equivalence based on fully matching resolutions is still strictly finer than the
corresponding ct-equivalences based on partially matching resolutions or extremal probabilities (the
counterexample provided by Fig.~\ref{fig:dis_vs_by} is still valid), it turns out that every ct-equivalence
based on partially matching resolutions coincides with the corresponding ct-equivalence based on extremal
probabilities. As far as $\sbis{\textrm{PTe-}\sqcup\sqcap}$ and $\sbis{\textrm{PTe-}\forall\exists}$ are
concerned, their ct-variants coincide as well. In other words, when moving to randomized schedulers, the
central fragment and the bottom fragment of the spectrum in Fig.~\ref{fig:spectrum_full} collapse.
Pictorially, all the ordinary arrows in Fig.~\ref{fig:spectrum_full} going from the central fragment to the
bottom one become bidirectional in the presence of randomized schedulers. Moreover, it holds that every
ct-equivalence based on extremal probabilities coincides with the corresponding equivalence in the bottom
fragment of the spectrum in Fig.~\ref{fig:spectrum_full}.

	\begin{theorem}\label{thm:rand_results}

It holds that:

		\begin{enumerate}

\item $\sbis{} \: \subseteq \: \sbis{}^{\rm ct}$ for every considered equivalence $\sbis{}$.

\item $\sbis{\rm \pi,dis}^{\rm ct} \: \subseteq \: \sbis{\rm \pi}^{\rm ct} \: = \: \sbis{\rm
\pi,\sqcup\sqcap}^{\rm ct} \: = \: \sbis{\rm \pi,\sqcup\sqcap}$ for all $\pi \in \{ {\rm PB},
\textrm{PTe-tbt}, {\rm PRTr}, {\rm PFTr}, {\rm PR}, {\rm PF}, {\rm PCTr}, {\rm PTr} \}$ over image-finite
NPLTS models.

\item $\sbis{\rm \pi,dis}^{\rm ct} \: \subseteq \: \sbis{\rm \pi}^{\rm ct} \: = \: \sbis{\rm
\pi,\sqcup}^{\rm ct} \: = \: \sbis{\rm \pi,\sqcup}$ for all $\pi \in \{ {\rm PS}, {\rm PCS}, {\rm PFS}, {\rm
PRS} \}$ over image-finite NPLTS models.

\item $\sbis{\textrm{PTe-}\sqcup\sqcap} \: = \: \sbis{\textrm{PTe-}\sqcup\sqcap}^{\rm ct} \: = \:
\sbis{\textrm{PTe-}\forall\exists}^{\rm ct} \: = \: \sbis{\textrm{PTe-tbt,dis}}^{\rm ct}$ over image-finite
NPLTS models.

		\end{enumerate}

		\begin{proof}
Let $(S, A, \! \arrow{}{} \!)$ be an NPLTS and $s_{1}, s_{2} \in S$:

			\begin{enumerate}

\item Since matching ordinary transitions induce matching combined transitions, $\sbis{}^{\rm ct}$ performs
at least the same identifications as $\sbis{}$.

\item The proof of the fact that $\sbis{\rm \pi,dis}^{\rm ct} \: \subseteq \: \sbis{\rm \pi}^{\rm ct} \:
\subseteq \: \sbis{\rm \pi,\sqcup\sqcap}^{\rm ct}$ is similar to the proof of
Thm.~6.5(1) in~\cite{BDL13TCS} when $\pi = {\rm PB}$, to the proof of Thm.~5.9(2) in~\cite{BDL13TCS}
when $\pi = \textrm{PTe-tbt}$, and to the proof of Thm.~3.5 in~\cite{BDL13TCS} in all the other cases.
Moreover, it holds that $\sbis{\rm \pi,\sqcup\sqcap}^{\rm ct} \: \subseteq \: \sbis{\rm \pi}^{\rm ct}$ (and
hence $\sbis{\rm \pi}^{\rm ct} \: = \: \sbis{\rm \pi,\sqcup\sqcap}^{\rm ct}$) when the NPLTS is image
finite. \\
Consider the case $\pi = {\rm PB}$ and suppose that $s_{1} \sbis{\rm PB,\sqcup\sqcap}^{\rm ct} s_{2}$. This
means that there exists a \linebreak ct-probabilistic $\sqcup\sqcap$-bisimulation $\calb$ over $S$ such that
$(s_{1}, s_{2}) \in \calb$. Given $\calg \in 2^{S / \calb}$ and $a \in A$, assume that there exists $s_{1}
\arrow{a}{} \cald_{1}$ such that $\cald_{1}(\bigcup \calg) = p$. Since $(s_{1}, s_{2}) \in \calb$ and the
NPLTS is image finite, there exist $s_{2} \arrow{a}{\rm c} \cald'_{2}$ such that $\cald'_{2}(\bigcup \calg)
= p' \le p$ and $s_{2} \arrow{a}{\rm c} \cald''_{2}$ such that $\cald''_{2}(\bigcup \calg) = p'' \ge p$. If
$p' = p$ (resp.\ $p'' = p$), then $s_{1} \arrow{a}{} \cald_{1}$ is trivially matched by $s_{2} \arrow{a}{\rm
c} \cald'_{2}$ (resp.\ $s_{2} \arrow{a}{\rm c} \cald''_{2}$) with respect to $\sbis{\rm PB}^{\rm ct}$ when
considering $\calg$. Assume that $p' < p < p''$ and note that $s_{2} \arrow{a}{\rm c} (x \cdot \cald'_{2} +
y \cdot \cald''_{2})$ for all $x, y \in \realns_{]0, 1]}$ such that $x + y = 1$. Indeed, directly from the
definition of combined transition, we have that:

				\begin{itemize}

\item Since $s_{2} \arrow{a}{\rm c} \cald'_{2}$, there exist $n \in \natns_{> 0}$, $\{ p'_{i} \in
\realns_{]0, 1]} \mid 1 \le i \le n \}$, and $\{ s_{2} \arrow{a}{} \hat\cald'_{i} \mid 1 \le i \le n \}$
such that $\sum_{i = 1}^{n} p'_{i} = 1$ and $\sum_{i = 1}^{n} p'_{i} \cdot \hat\cald'_{i} = \cald'_{2}$.

\item Since $s_{2} \arrow{a}{\rm c} \cald''_{2}$, there exist $m \in \natns_{> 0}$, $\{ p''_{j} \in
\realns_{]0, 1]} \mid 1 \le j \le m \}$, and $\{ s_{2} \arrow{a}{} \hat\cald''_{j} \mid 1 \le j \le m \}$
such that $\sum_{j = 1}^{m} p''_{j} = 1$ and $\sum_{j = 1}^{m} p''_{j} \cdot \hat\cald''_{j} = \cald''_{2}$.

				\end{itemize}

Hence, $(x \cdot \cald'_{2} + y \cdot \cald''_{2})$ can be obtained from the appropriate combination of:
\cws{0}{\hspace*{-0.8cm} \{ s_{2} \arrow{a}{} \hat\cald'_{i} \mid 1 \le i \le n \} \cup \{ s_{2} \arrow{a}{}
\hat\cald''_{j} \mid 1 \le j \le m \}}
with coefficients:
\cws{0}{\hspace*{-0.8cm} \{ x \cdot p'_{i} \in \realns_{]0, 1]} \mid 1 \le i \le n \} \cup \{ y \cdot
p''_{j} \in \realns_{]0, 1]} \mid 1 \le j \le m \}}
If we take $x = \frac{p'' - p}{p'' - p'}$ and $y = \frac{p - p'}{p'' - p'}$, then $s_{2} \arrow{a}{\rm c}
\left( \frac{p'' - p}{p'' - p'} \cdot \cald'_{2} + \frac{p - p'}{p'' - p'} \cdot \cald''_{2} \right)$ with:
\cws{0}{\hspace*{-0.8cm}\begin{array}{rcl}
\left( \frac{p'' - p}{p'' - p'} \cdot \cald'_{2} + \frac{p - p'}{p'' - p'} \cdot \cald''_{2} \right)(\bigcup
\calg) & \!\!\! = \!\!\! & \frac{p'' - p}{p'' - p'} \cdot \cald'_{2}(\bigcup \calg) + \frac{p - p'}{p'' -
p'} \cdot \cald''_{2}(\bigcup \calg) \\
& \!\!\! = \!\!\! & \frac{p'' - p}{p'' - p'} \cdot p' + \frac{p - p'}{p'' - p'} \cdot p'' \: = \: \frac{p'
\cdot p'' - p \cdot p' + p \cdot p'' - p' \cdot p''}{p'' - p'} \\
& \!\!\! = \!\!\! & p \cdot \frac{p'' - p'}{p'' - p'} \: = \: p \: = \: \cald_{1}(\bigcup \calg) \\
\end{array}}
Due to the generality of $(s_{1}, s_{2}) \in \calb$, $a \in A$, and $\calg \in 2^{S / \calb}$, it turns out
that $\calb$ is also a ct-probabilistic bisimulation, i.e., $s_{1} \sbis{\rm PB}^{\rm ct} s_{2}$. \\
The proof for the other seven cases is similar, with actions being replaced by traces and transitions being
replaced by resolutions. For instance, suppose that $s_{1} \sbis{\rm PTr,\sqcup\sqcap}^{\rm ct} s_{2}$.
Given $\alpha \in A^{*}$, assume that there exists $\calz_{1} \in \ms{Res}_{\alpha}^{\rm ct}(s_{1})$ such
that $\ms{prob}(\calcc(z_{s_{1}}, \alpha)) = p$. Since $s_{1} \sbis{\rm PTr,\sqcup\sqcap}^{\rm ct} s_{2}$
and the NPLTS is image finite, there exist $\calz'_{2}, \calz''_{2} \in \ms{Res}_{\alpha}^{\rm ct}(s_{2})$
such that $\ms{prob}(\calcc(z'_{s_{2}}, \alpha)) = p' \le p$ and $\ms{prob}(\calcc(z''_{s_{2}}, \alpha)) =
p'' \ge p$. If $p' = p$ (resp.\ $p'' = p$), then $\calz_{1}$ is trivially matched by $\calz'_{2}$ (resp.\
$\calz''_{2}$) with respect to $\sbis{\rm PTr}^{\rm ct}$ when considering $\alpha$. Assume that $p' < p <
p''$ and consider the resolution $\calz_{2} = x \cdot \calz'_{2} + y \cdot \calz''_{2}$ of $s_{2}$ defined
as follows for $x, y \in \realns_{]0, 1]}$ such that $x + y = 1$. Since $p' \neq p''$ and they both refer to
the probability of performing an $\alpha$-compatible computation from $s_{2}$, the two resolutions
$\calz'_{2}$ and $\calz''_{2}$ of $s_{2}$ differ at least in one point in which the nondeterministic choice
between two transitions labeled with the same action occurring in $\alpha$ has been resolved differently. We
obtain $\calz_{2}$ from $\calz'_{2}$ and $\calz''_{2}$ by combining the two different transitions into a
single one with coefficients $x$ and $y$ for their target distributions, respectively, in the first of those
points. If we take $x = \frac{p'' - p}{p'' - p'}$ and $y = \frac{p - p'}{p'' - p'}$, then:
\cws{0}{\hspace*{-0.8cm}\begin{array}{rcl}
\ms{prob}(\calcc(z_{s_{2}}, \alpha)) & \!\!\! = \!\!\! & \frac{p'' - p}{p'' - p'} \cdot 
\ms{prob}(\calcc(z'_{s_{2}}, \alpha)) + \frac{p - p'}{p'' - p'} \cdot \ms{prob}(\calcc(z''_{s_{2}}, \alpha))
\\
& \!\!\! = \!\!\! & \frac{p'' - p}{p'' - p'} \cdot p' + \frac{p - p'}{p'' - p'} \cdot p'' \: = \: \frac{p'
\cdot p'' - p \cdot p' + p \cdot p'' - p' \cdot p''}{p'' - p'} \\
& \!\!\! = \!\!\! & p \cdot \frac{p'' - p'}{p'' - p'} \: = \: p \: = \: \ms{prob}(\calcc(z_{s_{1}}, \alpha))
\\
\end{array}}
Due to the generality of $\alpha \in A^{*}$, it turns out that $s_{1} \sbis{\rm PTr}^{\rm ct} s_{2}$. \\
The fact that $\sbis{\rm \pi,\sqcup\sqcap} \: \subseteq \: \sbis{\rm \pi,\sqcup\sqcap}^{\rm ct}$ stems from
the first result of this theorem. Moreover, it holds that $\sbis{\rm \pi,\sqcup\sqcap}^{\rm ct} \: \subseteq
\: \sbis{\rm \pi,\sqcup\sqcap}$ (and hence $\sbis{\rm \pi,\sqcup\sqcap}^{\rm ct} \: = \: \sbis{\rm
\pi,\sqcup\sqcap}$) when the NPLTS is image finite. \\
Consider the case $\pi = {\rm PB}$ and suppose that $s_{1} \sbis{\rm PB,\sqcup\sqcap}^{\rm ct} s_{2}$. This
means that there exists a \linebreak ct-probabilistic $\sqcup\sqcap$-bisimulation $\calb$ over $S$ such that
$(s_{1}, s_{2}) \in \calb$. In other words, whenever $(s'_{1}, s'_{2}) \in \calb$, then for all $\calg \in
2^{S / \calb}$ and $a \in A$ it holds that $s'_{1} \arrow{a}{} \!$ iff $s'_{2} \arrow{a}{} \!$ and:
\cws{0}{\hspace*{-0.8cm}\begin{array}{rcl}
\bigsqcup\limits_{s'_{1} \arrow{a}{\rm c} \cald_{1}} \cald_{1}(\bigcup \calg) & \!\!\! = \!\!\! &
\bigsqcup\limits_{s'_{2} \arrow{a}{\rm c} \cald_{2}} \cald_{2}(\bigcup \calg) \\[0.4cm]
\bigsqcap\limits_{s'_{1} \arrow{a}{\rm c} \cald_{1}} \cald_{1}(\bigcup \calg) & \!\!\! = \!\!\! &
\bigsqcap\limits_{s'_{2} \arrow{a}{\rm c} \cald_{2}} \cald_{2}(\bigcup \calg) \\
\end{array}}
Since the NPLTS is image finite, given $\calg \in 2^{S / \calb}$, $a \in A$, and $s \in S$ having at least
one outgoing \linebreak $a$-transition, it holds that:
\cws{0}{\hspace*{-0.8cm}\begin{array}{rcl}
\bigsqcup\limits_{s \arrow{a}{\rm c} \cald} \cald(\bigcup \calg) & \!\!\! = \!\!\! & \bigsqcup\limits_{s
\arrow{a}{} \cald} \cald(\bigcup \calg) \\[0.4cm]
\bigsqcap\limits_{s \arrow{a}{\rm c} \cald} \cald(\bigcup \calg) & \!\!\! = \!\!\! & \bigsqcap\limits_{s
\arrow{a}{} \cald} \cald(\bigcup \calg) \\
\end{array}}
because the supremum and the infimum on the left are respectively achieved by two ordinary \linebreak
$a$-transitions of $s$. In fact, let $\cald_{\sqcup}$ (resp.\ $\cald_{\sqcap}$) be the target of an
$a$-transition of $s$ assigning the maximum (resp.\ minimum) value to $\bigcup \calg$ among all the
$a$-transitions of $s$ and consider an arbitrary convex combination of a subset $\{ s \arrow{a}{} \cald_{i}
\mid 1 \le i \le n \}$ of those transitions, with coefficients $p_{1}, \dots, p_{n}$ and $n \in \natns_{>
0}$. Then:
\cws{0}{\hspace*{-0.8cm}\begin{array}{rcccl}
\sum\limits_{i = 1}^{n} p_{i} \cdot \cald_{i}(\bigcup \calg) & \!\!\! \le \!\!\! & \sum\limits_{i = 1}^{n}
p_{i} \cdot \cald_{\sqcup}(\bigcup \calg) & \!\!\! = \!\!\! & \cald_{\sqcup}(\bigcup \calg) \\[0.3cm]
\sum\limits_{i = 1}^{n} p_{i} \cdot \cald_{i}(\bigcup \calg) & \!\!\! \ge \!\!\! & \sum\limits_{i = 1}^{n}
p_{i} \cdot \cald_{\sqcap}(\bigcup \calg) & \!\!\! = \!\!\! & \cald_{\sqcap}(\bigcup \calg) \\
\end{array}}
As a consequence, whenever $(s'_{1}, s'_{2}) \in \calb$, then for all $a \in A$ and $\calg \in 2^{S /
\calb}$ it holds that $s'_{1} \arrow{a}{} \!$ iff $s'_{2} \arrow{a}{} \!$ and:
\cws{0}{\hspace*{-0.8cm}\begin{array}{rcl}
\bigsqcup\limits_{s'_{1} \arrow{a}{} \cald_{1}} \cald_{1}(\bigcup \calg) & \!\!\! = \!\!\! &
\bigsqcup\limits_{s'_{2} \arrow{a}{} \cald_{2}} \cald_{2}(\bigcup \calg) \\[0.4cm]
\bigsqcap\limits_{s'_{1} \arrow{a}{} \cald_{1}} \cald_{1}(\bigcup \calg) & \!\!\! = \!\!\! &
\bigsqcap\limits_{s'_{2} \arrow{a}{} \cald_{2}} \cald_{2}(\bigcup \calg) \\
\end{array}}
This means that $\calb$ is also a probabilistic $\sqcup\sqcap$-bisimulation, i.e., $s_{1} \sbis{\rm
PB,\sqcup\sqcap} s_{2}$. \\
The proof for the other seven cases is similar, with actions being replaced by traces and transitions being
replaced by resolutions. For instance, suppose that $s_{1} \sbis{\rm PTr,\sqcup\sqcap}^{\rm ct} s_{2}$. This
means that for all $\alpha \in A^{*}$:
\cws{0}{\hspace*{-0.8cm}\begin{array}{rcl}
\bigsqcup\limits_{\calz_{1} \in \ms{Res}^{\rm ct}_{\alpha}(s_{1})} \ms{prob}(\calcc(z_{s_{1}}, \alpha)) &
\!\!\! = \!\!\! & \bigsqcup\limits_{\calz_{2} \in \ms{Res}^{\rm ct}_{\alpha}(s_{2})}
\ms{prob}(\calcc(z_{s_{2}}, \alpha)) \\[0.4cm]
\bigsqcap\limits_{\calz_{1} \in \ms{Res}^{\rm ct}_{\alpha}(s_{1})} \ms{prob}(\calcc(z_{s_{1}}, \alpha)) &
\!\!\! = \!\!\! & \bigsqcap\limits_{\calz_{2} \in \ms{Res}^{\rm ct}_{\alpha}(s_{2})}
\ms{prob}(\calcc(z_{s_{2}}, \alpha)) \\
\end{array}}
Given $\alpha \in A^{*}$ and $s \in S$, it holds that:
\cws{0}{\hspace*{-0.8cm}\begin{array}{rcl}
\bigsqcup\limits_{\calz \in \ms{Res}^{\rm ct}_{\alpha}(s)} \ms{prob}(\calcc(z_{s}, \alpha)) & \!\!\! =
\!\!\! & \bigsqcup\limits_{\calz \in \ms{Res}_{\alpha}(s)} \ms{prob}(\calcc(z_{s}, \alpha)) \\[0.4cm]
\bigsqcap\limits_{\calz \in \ms{Res}^{\rm ct}_{\alpha}(s)} \ms{prob}(\calcc(z_{s}, \alpha)) & \!\!\! =
\!\!\! & \bigsqcap\limits_{\calz \in \ms{Res}_{\alpha}(s)} \ms{prob}(\calcc(z_{s}, \alpha)) \\
\end{array}}
In fact, observing that:
\cws{0}{\hspace*{-0.8cm} \bigsqcup\limits_{\calz \in \ms{Res}^{\rm ct}_{\alpha}(s)} \ms{prob}(\calcc(z_{s},
\alpha)) \: \ge \: \bigsqcup\limits_{\calz \in \ms{Res}_{\alpha}(s)} \ms{prob}(\calcc(z_{s}, \alpha))}
because the set of probabilities on the left contains the set of probabilities on the right (a dual property
based on $\le$ holds for infima), we prove that:
\cws{0}{\hspace*{-0.8cm} \bigsqcup\limits_{\calz \in \ms{Res}^{\rm ct}_{\alpha}(s)} \ms{prob}(\calcc(z_{s},
\alpha)) \: \le \: \bigsqcup\limits_{\calz \in \ms{Res}_{\alpha}(s)} \ms{prob}(\calcc(z_{s}, \alpha))}
by proceeding by induction on the length $n$ of $\alpha$ (a dual property based on $\ge$ can be proved for
infima):

				\begin{itemize}

\item If $n = 0$, i.e., $\alpha = \varepsilon$, then:
\cws{4}{\hspace*{-1.6cm} \bigsqcup\limits_{\calz \in \ms{Res}^{\rm ct}_{\alpha}(s)} \ms{prob}(\calcc(z_{s},
\alpha)) \: = \: 1 \: = \: \bigsqcup\limits_{\calz \in \ms{Res}_{\alpha}(s)} \ms{prob}(\calcc(z_{s},
\alpha))}

\item Let $n \in \natns_{> 0}$ and suppose that the property holds for all traces of length $m = 0, \dots, n
- 1$. Assume that $\alpha = a \, \alpha'$. If $s$ has no outgoing $a$-transitions (an outgoing
non-$a$-transition in the case of infima), then:
\cws{4}{\hspace*{-1.6cm} \bigsqcup\limits_{\calz \in \ms{Res}^{\rm ct}_{\alpha}(s)} \ms{prob}(\calcc(z_{s},
\alpha)) \: = \: 0 \: = \: \bigsqcup\limits_{\calz \in \ms{Res}_{\alpha}(s)} \ms{prob}(\calcc(z_{s},
\alpha))}
otherwise, indicating with $s \arrow{a}{\rm c} \cald_{\rm c}$ a combined transition from $s$ with
$\cald_{\rm c} = \sum_{i = 1}^{m} p_{i} \cdot \cald_{i}$, \linebreak we have that:
\cws{0}{\hspace*{-1.6cm}\begin{array}{l}
\bigsqcup\limits_{\calz \in \ms{Res}^{\rm ct}_{\alpha}(s)} \ms{prob}(\calcc(z_{s}, \alpha)) \: = \\
\hspace*{1.4cm} = \: \bigsqcup\limits_{s \arrow{a}{\rm c} \cald_{\rm c}} \, \sum\limits_{s' \in S} \left(
\cald_{\rm c}(s') \cdot \bigsqcup\limits_{\calz' \in \ms{Res}^{\rm ct}_{\alpha'}(s')}
\ms{prob}(\calcc(z'_{s'}, \alpha')) \right) \\
\hspace*{1.4cm} \le \: \bigsqcup\limits_{s \arrow{a}{\rm c} \cald_{\rm c}} \, \sum\limits_{s' \in S} \left(
\cald_{\rm c}(s') \cdot \bigsqcup\limits_{\calz' \in \ms{Res}_{\alpha'}(s')} \ms{prob}(\calcc(z'_{s'},
\alpha')) \right) \\
\hspace*{1.4cm} = \: \bigsqcup\limits_{s \arrow{a}{\rm c} \cald_{\rm c}} \, \sum\limits_{s' \in S} \left(
\sum\limits_{i = 1}^{m} (p_{i} \cdot \cald_{i}(s')) \cdot \bigsqcup\limits_{\calz' \in
\ms{Res}_{\alpha'}(s')} \ms{prob}(\calcc(z'_{s'}, \alpha')) \right) \\
\hspace*{1.4cm} = \: \bigsqcup\limits_{s \arrow{a}{\rm c} \cald_{\rm c}} \, \sum\limits_{i = 1}^{m} p_{i}
\cdot \left( \sum\limits_{s' \in S} \left( \cald_{i}(s') \cdot \bigsqcup\limits_{\calz' \in
\ms{Res}_{\alpha'}(s')} \ms{prob}(\calcc(z'_{s'}, \alpha')) \right) \right) \\
\hspace*{1.4cm} \le \: \bigsqcup\limits_{s \arrow{a}{\rm c} \cald_{\rm c}} \, \sum\limits_{i = 1}^{m} p_{i}
\cdot \bigsqcup\limits_{i = 1}^{m} \left( \sum\limits_{s' \in S} \left( \cald_{i}(s') \cdot
\bigsqcup\limits_{\calz' \in \ms{Res}_{\alpha'}(s')} \ms{prob}(\calcc(z'_{s'}, \alpha')) \right) \right) \\
\hspace*{1.4cm} = \: \bigsqcup\limits_{s \arrow{a}{\rm c} \cald_{\rm c}} \, \bigsqcup\limits_{i = 1}^{m}
\left( \sum\limits_{s' \in S} \left( \cald_{i}(s') \cdot \bigsqcup\limits_{\calz' \in
\ms{Res}_{\alpha'}(s')} \ms{prob}(\calcc(z'_{s'}, \alpha')) \right) \right) \\
\hspace*{1.4cm} = \: \bigsqcup\limits_{s \arrow{a}{} \cald} \, \sum\limits_{s' \in S} \left( \cald(s') \cdot
\bigsqcup\limits_{\calz' \in \ms{Res}_{\alpha'}(s')} \ms{prob}(\calcc(z'_{s'}, \alpha')) \right) \\[0.6cm]
\hspace*{1.4cm} = \: \bigsqcup\limits_{\calz \in \ms{Res}_{\alpha}(s)} \ms{prob}(\calcc(z_{s}, \alpha)) \\
\end{array}}
where in the third line we have exploited the induction hypothesis and in the seventh line the fact that
$\sum_{i = 1}^{m} p_{i} = 1$.

				\end{itemize}

As a consequence, for all $\alpha \in A^{*}$:
\cws{0}{\hspace*{-0.8cm}\begin{array}{rcl}
\bigsqcup\limits_{\calz_{1} \in \ms{Res}_{\alpha}(s_{1})} \ms{prob}(\calcc(z_{s_{1}}, \alpha)) & \!\!\! =
\!\!\! & \bigsqcup\limits_{\calz_{2} \in \ms{Res}_{\alpha}(s_{2})} \ms{prob}(\calcc(z_{s_{2}}, \alpha))
\\[0.4cm]
\bigsqcap\limits_{\calz_{1} \in \ms{Res}_{\alpha}(s_{1})} \ms{prob}(\calcc(z_{s_{1}}, \alpha)) & \!\!\! =
\!\!\! & \bigsqcap\limits_{\calz_{2} \in \ms{Res}_{\alpha}(s_{2})} \ms{prob}(\calcc(z_{s_{2}}, \alpha)) \\
\end{array}}
This means that $s_{1} \sbis{\rm PTr,\sqcup\sqcap} s_{2}$.

\item The proof of the fact that $\sbis{\rm \pi,dis}^{\rm ct} \: \subseteq \: \sbis{\rm \pi}^{\rm ct} \: =
\: \sbis{\rm \pi,\sqcup}^{\rm ct}$ is similar to the proof of Thm.~\ref{thm:sim_results}(1). \\
The proof of the fact that $\sbis{\rm \pi,\sqcup}^{\rm ct} \: = \: \sbis{\rm \pi,\sqcup}$ is similar to the
proof of the corresponding part of the second result of this theorem for bisimulation semantics.

\item We start by proving that $\sbis{\textrm{PTe-}\sqcup\sqcap} \: = \:
\sbis{\textrm{PTe-}\sqcup\sqcap}^{\rm ct}$. Given an arbitrary state $s \in S$ and an arbitrary NPT $\calt =
(O, A, \! \arrow{}{\calt} \!)$ with initial state $o \in O$, it holds that:
\cws{0}{\hspace*{-0.8cm}\begin{array}{rcl}
\bigsqcup\limits_{\calz \in \ms{Res}^{\rm ct}_{\rm max}(s, o)} \ms{prob}(\calsc(z_{s, o})) & \!\!\! = \!\!\!
& \bigsqcup\limits_{\calz \in \ms{Res}_{\rm max}(s, o)} \ms{prob}(\calsc(z_{s, o})) \\[0.4cm]
\bigsqcap\limits_{\calz \in \ms{Res}^{\rm ct}_{\rm max}(s, o)} \ms{prob}(\calsc(z_{s, o})) & \!\!\! = \!\!\!
& \bigsqcap\limits_{\calz \in \ms{Res}_{\rm max}(s, o)} \ms{prob}(\calsc(z_{s, o})) \\
\end{array}}
In fact, first of all we note that:
\cws{0}{\hspace*{-0.8cm} \bigsqcup\limits_{\calz \in \ms{Res}^{\rm ct}_{\rm max}(s, o)}
\ms{prob}(\calsc(z_{s, o})) \: \ge \: \bigsqcup\limits_{\calz \in \ms{Res}_{\rm max}(s, o)}
\ms{prob}(\calsc(z_{s, o}))}
because a deterministic scheduler is a special case of randomized scheduler and hence the set of
probabilities on the left contains the set of probabilities on the right (a dual property based on $\le$
holds for infima). Therefore, it suffices to show that:
\cws{0}{\hspace*{-0.8cm} \bigsqcup\limits_{\calz \in \ms{Res}^{\rm ct}_{\rm max}(s, o)}
\ms{prob}(\calsc(z_{s, o})) \: \le \: \bigsqcup\limits_{\calz \in \ms{Res}_{\rm max}(s, o)}
\ms{prob}(\calsc(z_{s, o}))}
as we prove below by proceeding by induction on the length $n$ of the longest successful computation from
$(s, o)$, which is finite because $\calt$ is finite (a dual property based on $\ge$ can be established for
infima):

				\begin{itemize}

\item If $n = 0$, i.e., $o = \omega$, then:
\cws{4}{\hspace*{-1.6cm} \bigsqcup\limits_{\calz \in \ms{Res}^{\rm ct}_{\rm max}(s, o)}
\ms{prob}(\calsc(z_{s, o})) \: = \: 1 \: = \: \bigsqcup\limits_{\calz \in \ms{Res}_{\rm max}(s, o)}
\ms{prob}(\calsc(z_{s, o}))}

\item Let $n \in \natns_{> 0}$ and suppose that the property holds for all configurations from which the
longest successful computation has length $m = 0, \dots, n - 1$. Indicating with $(s, o) \arrow{a}{\rm c}
\cald_{\rm c}$ a combined transition from $(s, o)$ with $\cald_{\rm c} = \sum_{i = 1}^{m} p_{i} \cdot
\cald_{i}$, we have that:
\cws{0}{\hspace*{-1.6cm}\begin{array}{l}
\hspace*{-0.2cm} \bigsqcup\limits_{\calz \in \ms{Res}^{\rm ct}_{\rm max}(s, o)} \hspace{-0.4cm}
\ms{prob}(\calsc(z_{s, o})) \: = \\
\hspace*{0.8cm} = \: \bigsqcup\limits_{(s, o) \arrow{a}{\rm c} \cald_{\rm c}} \, \sum\limits_{(s', o') \in S
\times O} \left( \cald_{\rm c}(s', o') \cdot \bigsqcup\limits_{\calz' \in \ms{Res}^{\rm ct}_{\rm max}(s',
o')} \hspace{-0.6cm} \ms{prob}(\calsc(z'_{s', o'})) \right) \\
\hspace*{0.8cm} \le \: \bigsqcup\limits_{(s, o) \arrow{a}{\rm c} \cald_{\rm c}} \, \sum\limits_{(s', o') \in
S \times O} \left( \cald_{\rm c}(s', o') \cdot \bigsqcup\limits_{\calz' \in \ms{Res}_{\rm max}(s', o')}
\hspace{-0.6cm} \ms{prob}(\calsc(z'_{s', o'})) \right) \\
\hspace*{0.8cm} = \: \bigsqcup\limits_{(s, o) \arrow{a}{\rm c} \cald_{\rm c}} \, \sum\limits_{(s', o') \in S
\times O} \left( \sum\limits_{i = 1}^{m} (p_{i} \cdot \cald_{i}(s', o')) \cdot \bigsqcup\limits_{\calz' \in
\ms{Res}_{\rm max}(s', o')} \hspace{-0.6cm} \ms{prob}(\calsc(z'_{s', o'})) \right) \\
\hspace*{0.8cm} = \: \bigsqcup\limits_{(s, o) \arrow{a}{\rm c} \cald_{\rm c}} \, \sum\limits_{i = 1}^{m}
p_{i} \cdot \left( \sum\limits_{(s', o') \in S \times O} \left( \cald_{i}(s', o') \cdot
\bigsqcup\limits_{\calz' \in \ms{Res}_{\rm max}(s', o')} \hspace{-0.6cm} \ms{prob}(\calsc(z'_{s', o'}))
\right) \right) \\
\hspace*{0.8cm} \le \: \bigsqcup\limits_{(s, o) \arrow{a}{\rm c} \cald_{\rm c}} \, \sum\limits_{i = 1}^{m}
p_{i} \cdot \bigsqcup\limits_{i = 1}^{m} \left( \sum\limits_{(s', o') \in S \times O} \left( \cald_{i}(s',
o') \cdot \bigsqcup\limits_{\calz' \in \ms{Res}_{\rm max}(s', o')} \hspace{-0.6cm} \ms{prob}(\calsc(z'_{s',
o'})) \right) \right) \\
\hspace*{0.8cm} = \: \bigsqcup\limits_{(s, o) \arrow{a}{\rm c} \cald_{\rm c}} \, \bigsqcup\limits_{i =
1}^{m} \left( \sum\limits_{(s', o') \in S \times O} \left( \cald_{i}(s', o') \cdot \bigsqcup\limits_{\calz'
\in \ms{Res}_{\rm max}(s', o')} \hspace{-0.6cm} \ms{prob}(\calsc(z'_{s', o'})) \right) \right) \\
\hspace*{0.8cm} = \: \bigsqcup\limits_{(s, o) \arrow{a}{} \cald} \, \sum\limits_{(s', o') \in S \times O}
\left( \cald(s', o') \cdot \bigsqcup\limits_{\calz' \in \ms{Res}_{\rm max}(s', o')} \hspace{-0.6cm}
\ms{prob}(\calsc(z'_{s', o'})) \right) \\[0.6cm]
\hspace*{0.8cm} = \: \bigsqcup\limits_{\calz \in \ms{Res}_{\rm max}(s, o)} \hspace{-0.4cm}
\ms{prob}(\calsc(z_{s, o})) \\
\end{array}}
where in the third line we have exploited the induction hypothesis and in the seventh line the fact that
$\sum_{i = 1}^{m} p_{i} = 1$.

				\end{itemize}

We now prove that $\sbis{\textrm{PTe-}\sqcup\sqcap} \: = \: \sbis{\textrm{PTe-}\forall\exists}^{\rm ct}$.
Suppose that $s_{1} \sbis{\textrm{PTe-}\sqcup\sqcap} s_{2}$ and consider an arbitrary NPT $\calt = (O, A, \!
\arrow{}{\calt} \!)$ with initial state $o \in O$, so that:
\cws{0}{\hspace*{-0.8cm}\begin{array}{rcccl}
\bigsqcup\limits_{\calz_{1} \in \ms{Res}_{\rm max}(s_{1}, o)} \ms{prob}(\calsc(z_{s_{1}, o})) & \!\!\! =
\!\!\! & p_{\sqcup} & \!\!\! = \!\!\! & \bigsqcup\limits_{\calz_{2} \in \ms{Res}_{\rm max}(s_{2}, o)}
\ms{prob}(\calsc(z_{s_{2}, o})) \\[0.4cm]
\bigsqcap\limits_{\calz_{1} \in \ms{Res}_{\rm max}(s_{1}, o)} \ms{prob}(\calsc(z_{s_{1}, o})) & \!\!\! =
\!\!\! & p_{\sqcap} & \!\!\! = \!\!\! & \bigsqcap\limits_{\calz_{2} \in \ms{Res}_{\rm max}(s_{2}, o)}
\ms{prob}(\calsc(z_{s_{2}, o})) \\
\end{array}}
If $p_{\sqcup} = p_{\sqcap}$, then all the maximal resolutions of $(s_{1}, o)$ and $(s_{2}, o)$ have the
same success probability, from which it trivially follows that $s_{1} \sbis{\textrm{PTe-}\forall\exists}
s_{2}$ and hence $s_{1} \sbis{\textrm{PTe-}\forall\exists}^{\rm ct} s_{2}$. \\
Recalling that the NPLTS is image finite and the test is finite so that $\ms{Res}_{\rm max}(s_{1}, o)$ and
$\ms{Res}_{\rm max}(s_{2}, o)$ are both finite, if $p_{\sqcup} > p_{\sqcap}$, then $p_{\sqcup}$ must be
achieved on $\calz_{1, \sqcup} \in \ms{Res}_{\rm max}(s_{1}, o)$ and $\calz_{2, \sqcup} \in \ms{Res}_{\rm
max}(s_{2}, o)$ exhibiting the same successful traces, otherwise -- observing that both resolutions must
have at least one successful trace, otherwise it would be $p_{\sqcup} = 0$ thus violating $p_{\sqcup} >
p_{\sqcap}$ -- states $s_{1}$ and $s_{2}$ would be distinguished with respect to
$\sbis{\textrm{PTe-}\sqcup\sqcap}$ by a test obtained from $\calt$ by making success reachable only along
the successful traces of the one of $\calz_{1, \sqcup}$ and $\calz_{2, \sqcup}$ having a successful trace
not possessed by the other, unless that resolution also contains all the successful traces of the other
resolution, in which case success must be made reachable only along the successful traces of the other
resolution in order to contradict $s_{1} \sbis{\textrm{PTe-}\sqcup\sqcap} s_{2}$. \\
Likewise, $p_{\sqcap}$ must be achieved on $\calz_{1, \sqcap} \in \ms{Res}_{\rm max}(s_{1}, o)$ and
$\calz_{2, \sqcap} \in \ms{Res}_{\rm max}(s_{2}, o)$ exhibiting the same unsuccessful maximal traces,
otherwise -- observing that both resolutions must have at least one unsuccessful maximal trace, otherwise it
would be $p_{\sqcap} = 1$ thus violating $p_{\sqcup} > p_{\sqcap}$ -- states $s_{1}$ and $s_{2}$ would be
distinguished with respect to $\sbis{\textrm{PTe-}\sqcup\sqcap}$ by a test obtained from $\calt$ by making
success reachable also along an unsuccessful maximal trace occurring only in either $\calz_{1, \sqcap}$ or
$\calz_{2, \sqcap}$. \\
By reasoning on the dual test $\calt'$ in which the final states of $\calt$ that are successful (resp.\
unsuccessful) are made unsuccessful (resp.\ successful), it turns out that $\calz_{1, \sqcup}$ and
$\calz_{2, \sqcup}$ must also exhibit the same unsuccessful maximal traces and that $\calz_{1, \sqcap}$ and
$\calz_{2, \sqcap}$ must also exhibit the same successful traces. \\
If $\calz_{1, \sqcup}$ and $\calz_{2, \sqcup}$ do not have sequences of initial transitions in common with
$\calz_{1, \sqcap}$ and $\calz_{2, \sqcap}$, \linebreak then $\calz_{1, \sqcup}$ and $\calz_{1, \sqcap}$ on
one side and $\calz_{2, \sqcup}$ and $\calz_{2, \sqcap}$ on the other side cannot generate via convex
combinations any new resolution that would arise from a randomized scheduler, otherwise they can generate
all such resolutions having a certain sequence of initial transitions, thus covering all the intermediate
success probabilities between $p_{\sqcup}$ and $p_{\sqcap}$ for that sequence of initial transitions. This
shows that for each $\calz_{1} \in \ms{Res}^{\rm ct}_{\rm max}(s_{1}, o)$ with that sequence of initial
transitions there exists $\calz_{2} \in \ms{Res}^{\rm ct}_{\rm max}(s_{2}, o)$ with that sequence of initial
transitions such that $\ms{prob}(\calsc(z_{s_{1}, o})) = \ms{prob}(\calsc(z_{s_{2}, o}))$, and vice versa.
\\
The same procedure can now be applied to the remaining resolutions in $\ms{Res}_{\rm max}(s_{1}, o)$ and
$\ms{Res}_{\rm max}(s_{2}, o)$ that are not convex combinations of previously considered resolutions,
starting from those among the remaining resolutions on which the maximal and minimal success probabilities
are achieved. We can thus conclude that $s_{1} \sbis{\textrm{PTe-}\forall\exists}^{\rm ct} s_{2}$. \\
The fact that $s_{1} \sbis{\textrm{PTe-}\forall\exists}^{\rm ct} s_{2}$ implies $s_{1}
\sbis{\textrm{PTe-}\sqcup\sqcap} s_{2}$ follows from the fact that $s_{1}
\sbis{\textrm{PTe-}\forall\exists}^{\rm ct} s_{2}$ implies $s_{1} \sbis{\textrm{PTe-}\sqcup\sqcap}^{\rm ct}
s_{2}$ (the proof is similar to that of Thm.~5.9(1) in~\cite{BDL13TCS}) and from
$\sbis{\textrm{PTe-}\sqcup\sqcap}^{\rm ct} \: = \: \sbis{\textrm{PTe-}\sqcup\sqcap}$. \\
Finally, $\sbis{\textrm{PTe-}\forall\exists}^{\rm ct} \: = \: \sbis{\textrm{PTe-tbt,dis}}^{\rm ct}$ is a
straightforward consequence of Thm.~5.9(2) in~\cite{BDL13TCS}.
\fullbox

			\end{enumerate}

		\end{proof}

	\end{theorem}

\bigskip
\noindent
\textbf{Acknowledgment}: This work has been partially funded by EU Collaborative FET project no.~257414
\emph{Autonomic Service-Component Ensemble} (ASCENS) and by MIUR PRIN project \emph{Compositionality,
Interaction, Negotiation, Autonomicity for the Future ICT Society} (CINA).

\bibliographystyle{plain}
\bibliography{companion}

\end{document}